\documentclass{pinchcr}   

\usepackage{graphicx}        
\usepackage{amsmath}

\title{The jamming scenario--- an introduction and outlook}

\author{Andrea J. Liu}
\affiliation{Department of Physics and Astronomy, University of
Pennsylvania, Philadelphia, PA 19104, USA}

\author{Sidney R. Nagel}
\affiliation{James Franck Institute, The University of Chicago,
Chicago, IL 60637, USA}

\author{Wim van Saarloos}
\affiliation{Instituut-Lorentz, LION, Leiden University, P.O.Box
9506, 2300 RA Leiden, The Netherlands}

\author{Matthieu Wyart}
\affiliation{Physics Department, New York University, 4 Washington
Place, New York, NY 10003, USA}

\begin{document}

\maketitle

\preface
The jamming scenario of disordered media, formulated about 10 years ago, has in recent years been advanced 
by analyzing model systems of granular media. This has led to
various new concepts that are increasingly being explored in
in a variety of systems. This chapter contains an introductory review of these recent developments and provides an outlook on their applicability to 
different physical
systems and on future directions. The first part of the paper is devoted to an overview of the findings for model systems of frictionless spheres, focussing on the excess of low-frequency modes as the jamming point is approached. Particular attention is paid to a discussion of the cross-over frequency and length scales that govern this approach. We then discuss 
the effects of particle asphericity and static friction, the
applicability to bubble models for wet foams in which the friction
is dynamic, the dynamical arrest in colloids, and the implications
for molecular glasses.

\section{Introduction}
\label{jamsection:1}

\begin{figure}[t]
\centering
\includegraphics*[width=.7\textwidth]{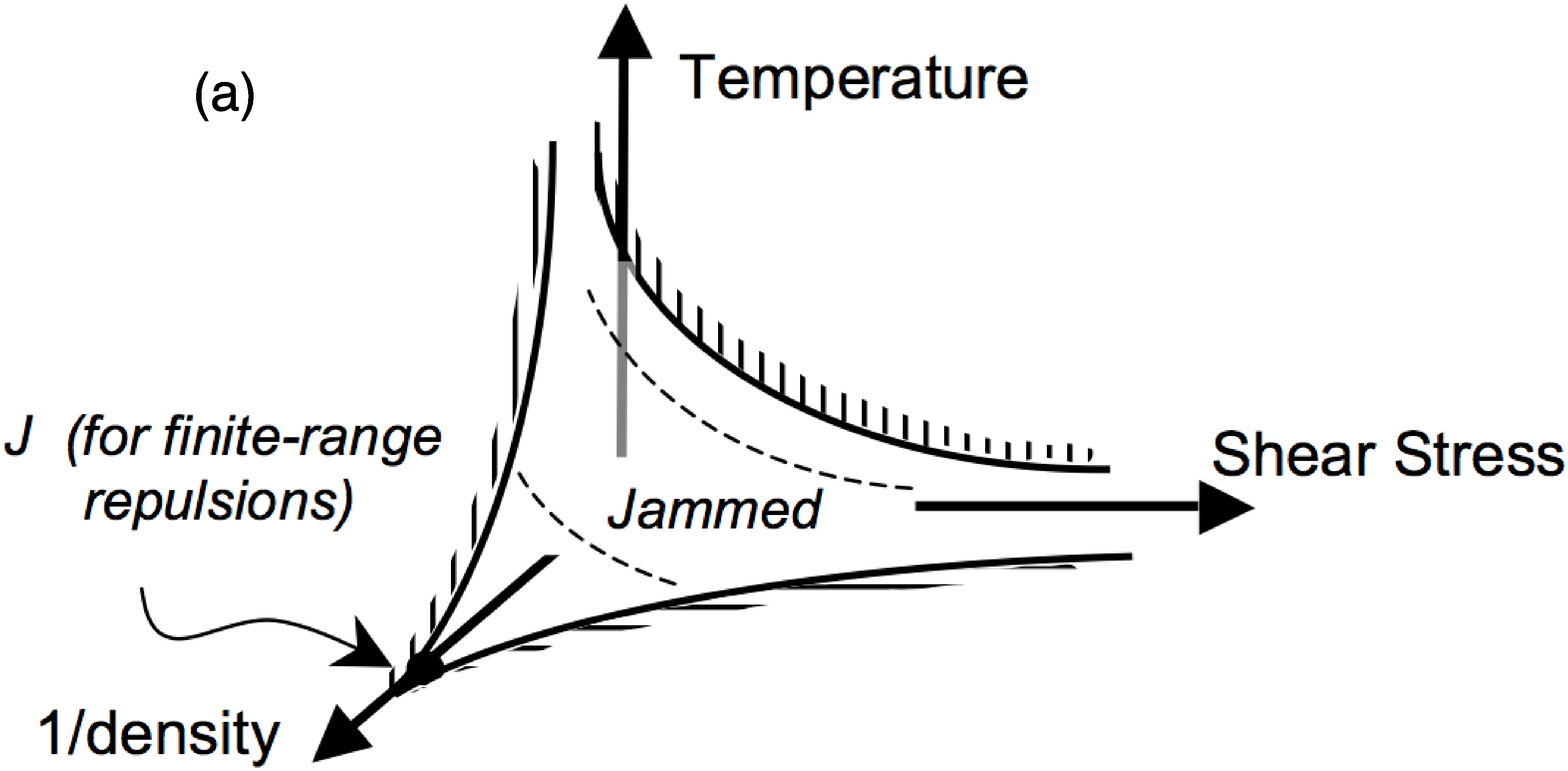}\hspace*{7mm}
\includegraphics*[width=.27\textwidth]{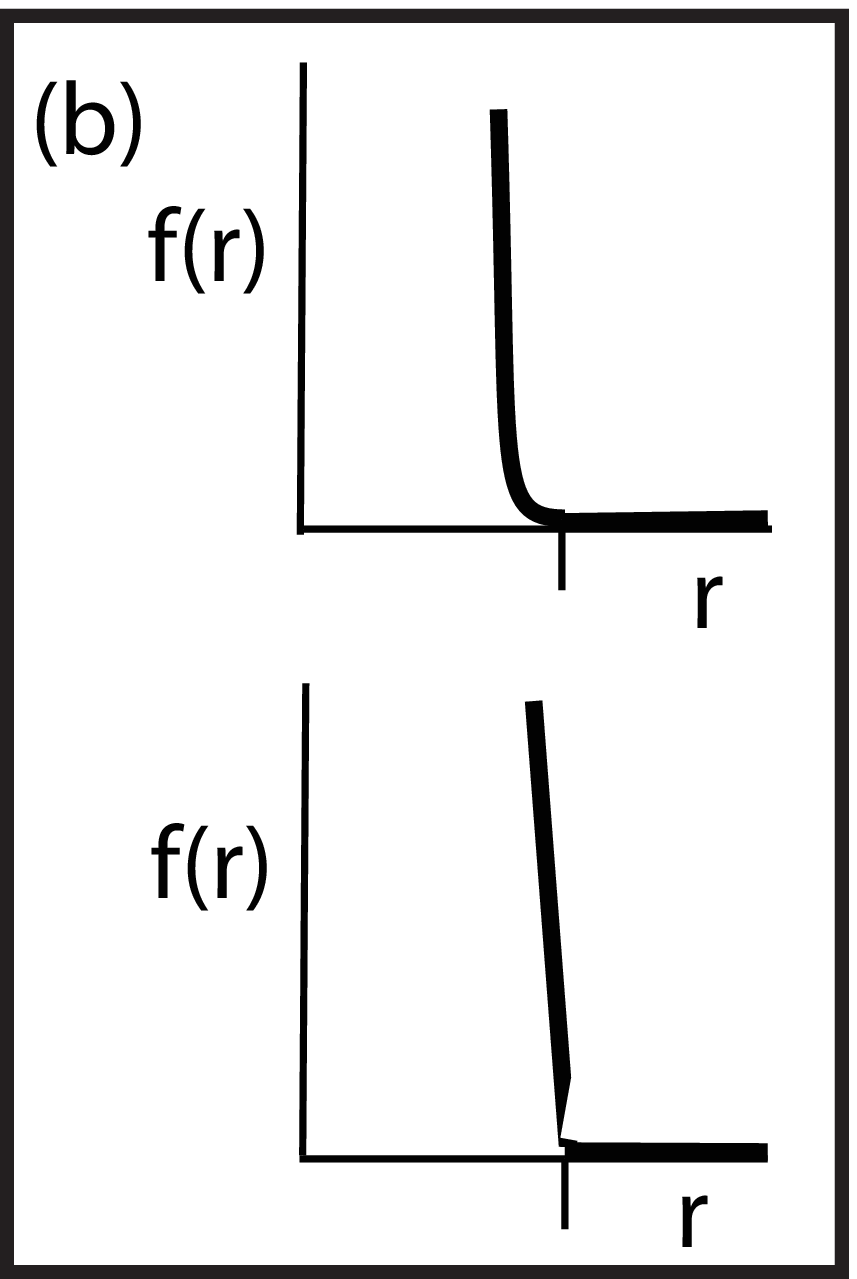}
\caption[]{(a) The jamming phase diagram; inside the shaded region,
for high density, low temperature and small shear stress,
particulate systems are {\em jammed}, {\em i.e.}, form a disordered
solid phase with a finite resistance to shear. The point $J$ is the
so-called ``jamming point'' for spherical particles with
finite-range repulsive forces. It acts like a critical point and
organizes the behavior in its neighborhood. (b) The type of forces
considered in models of frictionless spheres: the force $f(r)$ is of
finite range, {\em i.e.} vanishes when the distance $r$ between the
centers of the spheres exceeds some well-defined value. The lower
case, in which the force increases linearly with the compression
$\delta$, is often referred to as one-sided harmonic springs. The
Hertzian force of 3$d$ elastic spheres increases as $\delta^{3/2}$
and has a behavior as in the upper figure. } \label{jamfig:1}
\end{figure}

Just over ten years ago it was proposed \shortcite{jamnature} to approach disordered condensed-matter systems in a more unified way than is usually done, by starting from the observation that many systems --- not just molecular glasses, but also many soft-matter systems like granular media, colloids, pastes, emulsions, and foams --- exhibit a stiff solid phase at high density, provided the temperature 
and the external forces or stresses are small enough. This proposal
led to the idea of a {\em jamming phase diagram} for particulate
systems, redrawn in Fig.~\ref{jamfig:1}.

The very idea that there would be a generic jamming phase diagram is less trivial than 
may appear at first sight, since granular media, emulsions and foams
are in fact {\em athermal} systems --- they are represented by
points in the ground plane of Fig.~\ref{jamfig:1}, as their relevant
interaction energies are orders of magnitude larger than the thermal
energy scale $k_B T$.  While thermodynamics generally guarantees the
existence of a unique phase for molecular systems at equilibrium,
athermal systems strictly speaking lack the averaging needed to be
able to define a unique phase: their disordered jammed states are
often history dependent, a phenomenon they share with glasses.

Nevertheless, it has become clear that it is extremely useful to approach these systems with a unifying common focus in mind.  For particulate systems like granular media, the constituent particles essentially have  strong repulsive interactions of {\em finite} range: particles that are in contact do interact and repel each other, those that are not in contact, don't.  A detailed analysis of the simplest model of this type, that of $N$ repulsive frictionless spheres (or discs in two dimensions) in the large-$N$ limit, has revealed that 
the point at which they just jam into a rigid phase does indeed have
properties reminiscent of an asymptotic or critical point that
organizes the behavior in its neighborhood. This is the jamming
point, marked $J$ in Fig.~\ref{jamfig:1}. It is the aim of this
chapter to provide an introductory review of these recent findings,
and to provide an outlook on the relevance of our present
understanding for other systems. We aim our review of jamming on
those aspects that are of most interest from the point of view of
the main theme of this book, glasses. Hence we focus in particular
on the various length scales that emerge near the jamming point,
on the vibrational properties of the marginally jammed state above the jamming transition, 
 and on the connection of these findings with the dynamics.

This approach bears close similarities with investigations of the
rigidity of network glasses developed in the 1980's. Instead of
density and shear stress, Phillips argued that
the key control variable is the glass composition 
\shortcite{Phillips}.  
Based on a model of rigidity percolation proposed by Thorpe, where
springs are randomly deposited on a lattice
\shortcite{Thorpe,ThorpeBoolchand}, it was argued
that glass properties are controlled by a critical point at zero temperature.  
The jamming transition can be viewed as a special case of the
generic rigidity percolation problem in which the system
self-organized to avoid large fluctuations in its structure. Thus
the jamming of particles  is qualitatively different from  springs
randomly placed on a lattice.  In some respects, the jamming
transition is simpler, allowing the recent conceptual progress
reported here.


Unfortunately because of the focus of this book and space
limitations, this chapter cannot do justice to the wonderful recent
experimental developments: many new systems have become available or
have been analyzed in new ways, and are increasingly driving new
directions in the field of jamming. We refer for an overview of
recent experiments on colloids to the chapter by Cipelletti and
Weeks and for a review of experiments on grains and foams to the
chapter by Dauchot, Durian and van Hecke. We will point to relevant
experiments where appropriate, without going into details.

\section{Overview of recent result on jamming of frictionless sphere packings}

Close to the jamming point, packings of compressible frictionless
spheres exhibit various anomalies including a strongly enhanced
density of states at low frequencies. Of course, such
packings 
are a very idealized model system for granular media.  We postpone to section \ref{jamsection:3.0} a discussion of 
generalizations of this model such as the role of particle
anisotropy and friction.

In computer models of compressible spheres, one usually uses force laws of the type sketched in Fig.~\ref{jamfig:1}(b): two spheres of radii $R_i$ and $R_j$ experience zero force if the distance $r_{ij}$  between their centers is larger than $R_i+R_j$, and have a force that rapidly increases with the overlap $\delta_{ij}= R_i+R_j-r_{ij}$ if $r_{ij}<R_i+R_j$. A particularly convenient and common choice for computer models is the one-sided harmonic spring model, for which the repulsive force $f_{ij} $ for particles in contact increases linearly with the overlap  $\delta_{ij}$.  Another choice that one often encounters in the literature for compression of elastic balls, is the Hertzian force law $f_{ij} \sim \delta_{ij}^{3/2}$.  In studies of three-dimensional random packings, it usually suffices to take spheres whose radii are all the same, as one easily ends up with a random packing, but in two-dimensions one needs to take a polydisperse or bi-disperse distribution of discs in order to avoid crystallization. There are various ways to prepare static packings of spheres in mechanical equilibrium.  One method is to place particles randomly in a box at a fixed density and quench the system to its closest potential-energy minimum via conjugate-gradient or steepest descent algorithms \shortcite{epitome}. Other methods employ a slow inflation of all the radii
\footnote{This is somewhat like the Lubachevsky-Stillinger method to
generate random close packings of hard spheres
\shortcite{lubachevsky}.} to first generate a packing at the jamming
threshold where the majority of the particles experience minute
forces that balance on each particle.  This jamming threshold marks
the onset of a nonzero pressure and potential energy.  After this
initial preparation the packing is compressed (or dilated) while
continuously minimizing the energy \shortcite{epitome}.  Other
methods slowly adjust the radii so as to steer the pressure in the
packing to a prescribed value \shortcite{kostya}. We refer to the
literature for details.

We note that the onset packing fraction of the jamming transition
for an ensemble of states depends on the ensemble.   For systems
equilibrated at infinite temperature and quenched to $T=0$, the
onset packing fraction corresponds to that of an ensemble in which
each state, or local energy minimum, is weighted by the volume of
its energy basin.  In that case, $\phi_c \approx 0.64$ for
monodisperse spheres in three dimensions~\shortcite{epitome}.
However, if the ensemble is prepared by a quench from a system
equilibrated at a low temperature, the onset packing fraction is
higher~\shortcite{Witten,chaudhuri}.  It is clear from the
distribution of jamming onsets that this must be the
case~\shortcite{epitome}.  In the infinite-system-size limit, the
distribution approaches a delta-function at $\phi \approx 0.64$ but
with non-vanishing tails both on the high-density and low-density
side~\shortcite{epitome}.  In systems equilibrated at lower
temperatures, lower energy states are weighted more heavily.  Such
states have higher values of $\phi_c$, since the potential energy
increases with $\phi-\phi_c$.   Thus, the average value of $\phi_c$
for the ensemble must increase as the system is quenched from lower
temperatures, in accordance with the results of Chaudhuri, et
al.~\shortcite{chaudhuri}, and the system can be described as having
a line of jamming transitions extending upwards from $\phi \approx
0.64$, as in Ref.~\shortcite{Kurchan2}.  In all cases, however, the
onset pressure is zero, so the onset is sharp in terms of pressure
but not packing fraction, and the vibrational properties, etc. are
identical to those described below.

\subsection{The vibrational density of states of packings}

An important concept in condensed-matter systems is the density of states,  $D(\omega)$.  $D(\omega) {\rm d}\omega $ is proportional to the number of states with frequency between $\omega$ and $\omega+{\rm d}\omega$. Here $\omega$ refers to the frequency of the {\em vibrational normal modes} of the constituent particles. The density of states concept is also often used for for electronic states, but here we focus on the vibrational states of condensed matter systems. For a crystal, or in fact for any elastic medium,  $D(\omega)$ increases at low-frequencies as $\omega^{d-1}$, where $d$ is the dimension of space. This generic behavior arises from the fact that sound modes have a dispersion relation $\omega(k)$ which is linear in the wavenumber $k$, together with phase-space arguments for the number of modes with wavenumber between $k$ and $k+{\rm d}k$. The Debye scaling law $D(\omega) \sim \omega^2$ for $d=3$ underlies the ubiquitous $T^3$ low-temperature specific heat of three-dimensional solids. It serves as an important reference for identifying anomalous behavior --- {\em e.g.}, the well-known enhancement of the specific heat over the Debye law is an indication in glasses for an excess density of 
excitations at low frequencies.

\begin{figure}[t]
\centering
\includegraphics*[width=1.\textwidth]{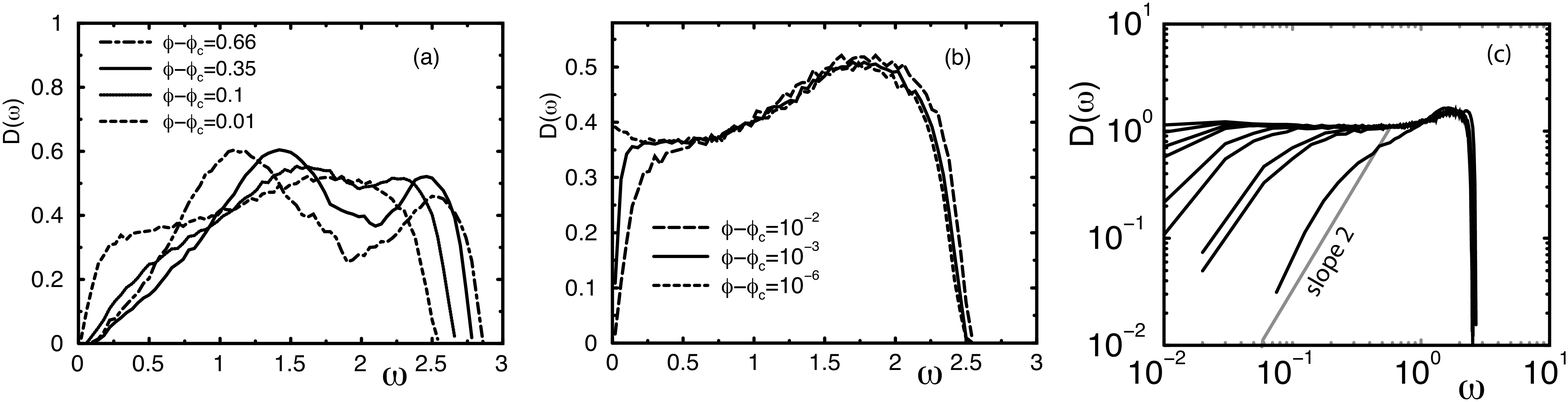}
\caption[]{The density of states of vibrational modes of
three-dimensional soft-sphere packings with one-sided harmonic
forces at varous densities: (a) significantly compressed samples;
(b) close to the jamming density $\phi_c$
\protect\shortcite{epitome}; (c) on logarithmic scale
\protect\shortcite{silbertPRL05}. } \label{jamfig:2}
\end{figure}

To analyze the vibrational modes, 
one obtains 
the so-called dynamical matrix familiar from solid state physics.
This dynamical matrix is essentially the second derivative of the
inter-particle potential, and hence has only non-zero elements for
particles that are in contact in the packing (for the one-sided
harmonic forces these terms are especially simple as the potential
is quadratic in the separation). From the diagonalization of the
dynamical matrix one then obtains the eigenmodes and their
eigenfrequencies $\omega$, and hence $D(\omega)$.

Fig.~\ref{jamfig:2} shows one of the early results for a
three-dimensional packing with one-sided harmonic forces
\shortcite{epitome,silbertPRL05}. Panel (a) shows that sufficiently
far above jamming, $D(\omega)$ vanishes at small frequencies, in
qualitative agreement with the Debye scenario, but as the density
$\phi$ is decreased towards the jamming density $\phi_c$, the weight
at small frequencies increases. Indeed, closer to the jamming
density $D(\omega)$ develops a plateau at small frequencies, see
panel (b). When plotted on a log-log scale, as in panel (c), the
crossover and the emergence of a plateau is even more clear: in
these data one observes the $\omega^2$ Debye scaling only for the
largest densities, while as the packings are decompressed towards
the jamming density, the plateau extends to lower and lower
frequencies. Clearly, the closer the packings are to the jamming
threshold, the more $D(\omega)$ is enhanced at low frequencies and
the larger are the deviations from the usual Debye behavior. We will
discuss in section \ref{section1.2.3} the cross-over frequency
$\omega^*$, that separates the plateau from the lower-frequency
downturn.

\subsection{Isostaticity and marginally connected solids}\label{sectionisostatic}
How do the excess low-frequency modes arise? The answer is related
to the fact that at the jamming threshold, packings of frictionless
discs and spheres are {\em isostatic}, {\em i.e.}, they are marginal
solids that can just maintain their stability
\shortcite{alexander,moukarzelPRL98,tkachenko1,tkachenko2,epitome,roux,wyartthesis}.
The origin of this is the following. Consider a disordered packing
with $N_c $ spheres or discs that have nontrivial
contacts\footnote{There is a subtlety here: in a typical packing
there is generally a small fraction of ``rattlers'' or ``floaters'',
particles in a large enough cage of other particles that, in the
absence of gravity, can float freely without any contact. These
should be left out from the counting below, and from the
determination of the average contact numer.}, with $Z$ their average
contact number. Force balance on each particle implies that the
vector sum of the forces adds up to zero; therefore the requirement
of force balance on all particles with contacts gives $dN_c$
conditions. If we view the contact forces between the particles as
the degrees of freedom that we have available to satisfy these
requirements, then there are $ZN_c/2$ such force degrees of freedom.
Clearly, assuming no special degeneracies, we arrive at the
condition
\begin{equation}
\mbox{for a stable packing:} ~~~~~Z \geq 2d   . \label{z>2d}
\end{equation}
in order that force balance can be maintained.

There is a second constraint in the limit that one approaches the
jamming point.  As all forces are purely repulsive, in this limit
almost all the individual contact forces must approach zero. For
force laws like those sketched in Fig.~\ref{jamfig:1}(b), this mean
that as one approaches the jamming point, all nontrivial $ZN_c/2$
contacts must obey the ``just-touching'' conditions $r_{ij} =
R_i+R_j$. The number of degrees of freedom associated with the
positions of the centers of the particles with contact is $dN_c$, so
if we think of putting the particles with contacts in the right
place to allow them to obey the just-touching conditions, we need to
have
\begin{equation}
\mbox{just-touching conditions at jamming:}~~~~~ Z \leq 2d.
\label{z<2d}
\end{equation}
Clearly, the value $Z_{\rm iso}=2d$, the isostatic value, has a
special significance. Indeed, the above two conditions imply that as
packings approach the jamming point, point J of
Fig.~\ref{jamfig:1}(a), from the jammed side, {\em e.g.}, by
decompressing them, one will have
\begin{equation}
\mbox{upon approaching point J:}~~~~~ Z \downarrow Z_{\rm iso}=2d.
\label{ziso}
\end{equation}
Note that these results are independent of the presence of
polydispersity and the details of the repulsive force law, provided
it is of finite range and continuous.\footnote{In passing, we note
that $\phi_c$, the density of a packing of monodisperse spheres (all
the same radii), approaches \shortcite{epitome} the random close
packing density of hard spheres.  There is an active line of
research aimed at analyzing and relating the concepts of jamming and
the maximally random jammed state, which has been proposed to
replace the concept of the random close packed state
\shortcite{torquato00,torquato01}.}

\begin{figure}[t]
\centering
\includegraphics*[width=.5\textwidth]{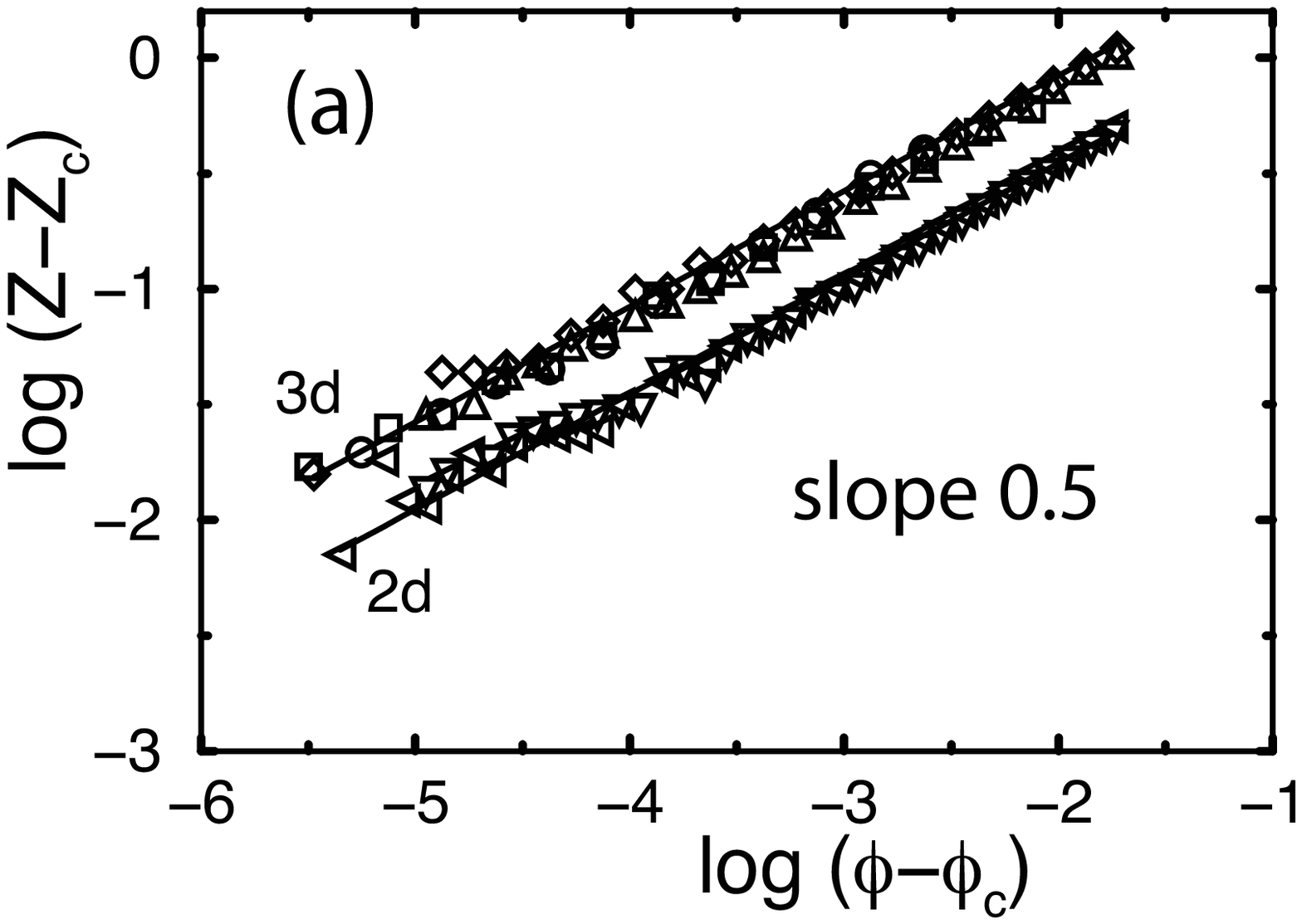}\hspace*{4mm}
\includegraphics*[width=.441\textwidth]{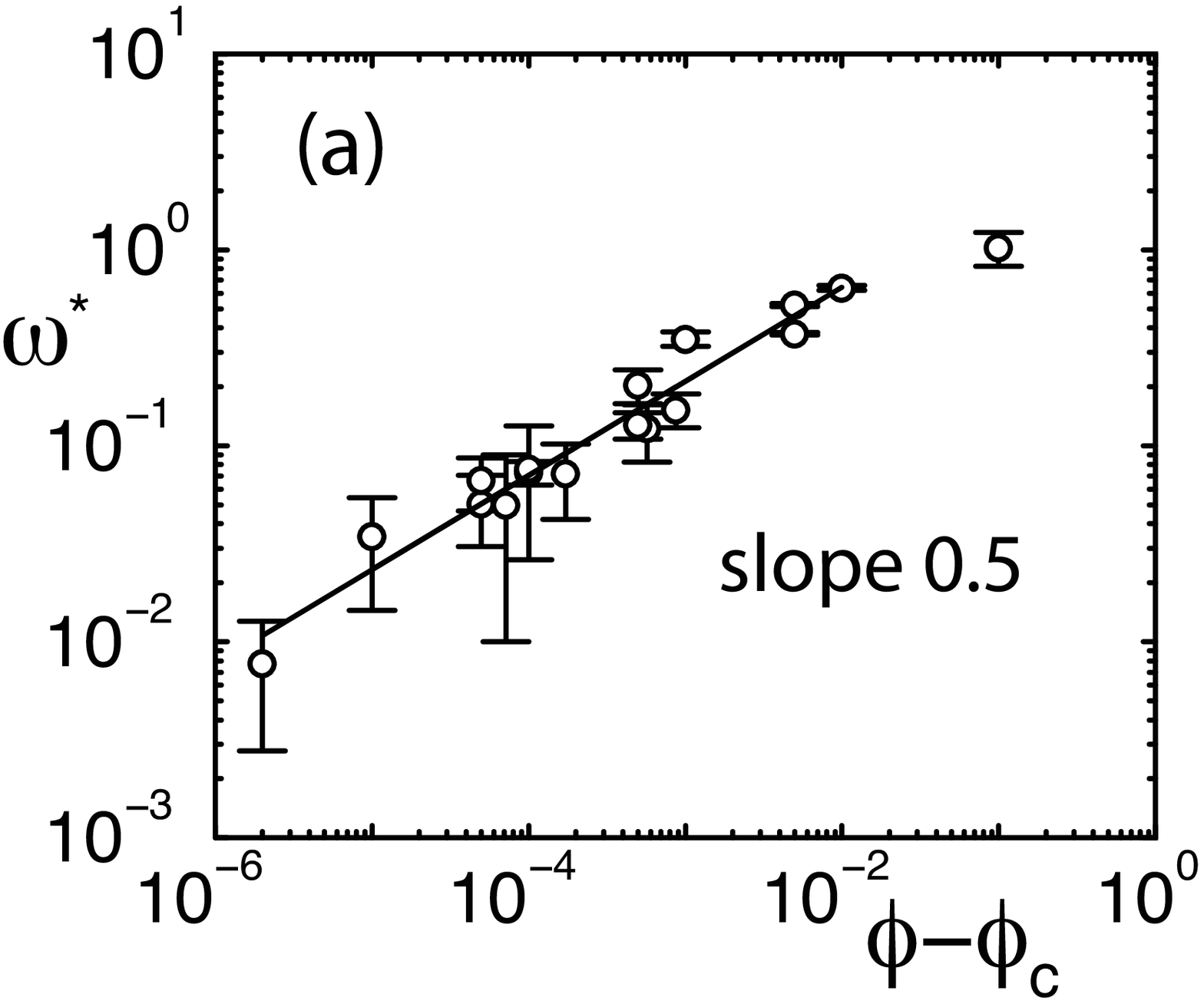}
\caption[]{(a) $Z$ scales as $|\Delta \phi|^{0.5}$, from
\protect\shortcite{epitome}. These results extend those of an
earlier numerical study~\protect\shortcite{durian95}.  (b)
$\omega^*$ scales as $|\Delta \phi|^{0.5}$. From
\protect\shortcite{silbertPRL05}. Together these two sets of data
are consistent with $\omega^* \sim \Delta Z$. } \label{jamfig:3}
\end{figure}

The above argument will be further justified in section
\ref{sectionmatt} for frictionless spheres.
Fig.~\ref{jamfig:3}(a) shows numerical simulation results for
$\Delta Z=Z-Z_{\rm iso}$, plotted on a log-log scale as a function
of the distance from jamming, $\Delta \phi=\phi-\phi_c$, in both 2
and 3 dimensions. In both cases $\Delta Z$ goes to zero at jamming:
$\Delta Z\sim \sqrt{\Delta \phi}$
~\shortcite{durian95,ohern02,epitome}. A recent experiment aimed at
testing this scaling, albeit in a system with friction, can be found
in \shortcite{behringer07}. The isostaticity concept will be
re-examinated in section \ref{jamsection:3.1} for ellipses.

Note that according to (\ref{z>2d}) the packings at isostaticity
have just enough contacts to maintain stability. In this sense, they
are marginal packings, packings at the edge of stability. Moreover,
if one imagines a packing with M fewer contacts than dictated by the
isostaticity condition, one can according to (\ref{z<2d}) deform the
packing in $M$ different directions in the space of coordinates,
while respecting the just-touching constraints (so that particles
are not pressed into each other). These therefore correspond to $M$
zero-energy deformation modes in which particles slide past one
another. These modes, which are in general global modes, have been
called floppy modes \shortcite{thorpe1983,alexander}.

\subsection{The plateau in the density of states and the cross-over frequency, $\omega^*$, and length scale, $\ell^*$.} \label{section1.2.3}

\begin{figure}[t]
\centering
\includegraphics*[width=0.99\textwidth]{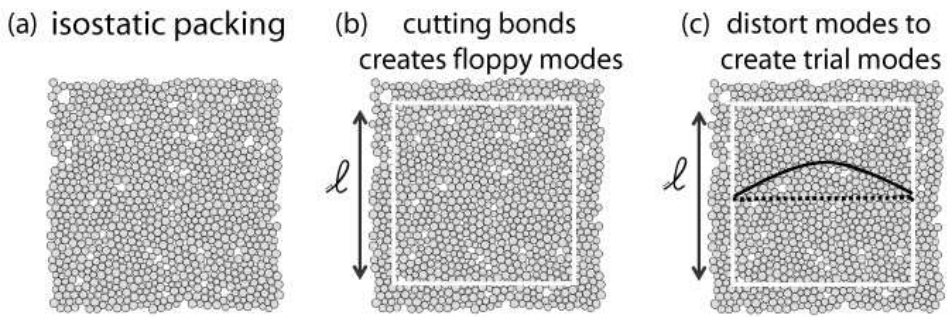}
\caption[]{Three stages of creating a low-energy Ansatz for a
vibrational eigenmode of the system, used in the argument to derive
the flatness of the $D(\omega)$ close to point J, as inspired by
\protect\shortcite{wyartEPL,wyartPRE}. In (c), the solid line
illustrates the behavior of the smooth amplitude with which the
underlying floppy mode is distorted, and which vanishes at the
boundaries of the box. See text for details.} \label{jamfig:4}
\end{figure}

The development of a plateau in the density of states is intimately
connected with the approach to the isostatic jamming point. To see
this, we argue as follows \shortcite{wyartEPL,wyartPRE}. Let us
start from an isostatic packing at jamming, sketched in
Fig.~\ref{jamfig:4}(a). Now imagine we disregard (``cut'') for a
moment the bonds across a square or cube of linear size $\ell$, as
sketched in panel (b). There are of order $\ell^{d-1}$ of these
surface bonds, and since the orginal packing was isostatic, by
cutting the bonds at the surface we create of order $\ell^{d-1}$
floppy zero-energy modes within the cube. Each of these modes will
be very complicated and disordered, but that does not matter here.
Next, as indicated in panel (c), we use each of these floppy modes
to create a variational Ansatz for a vibrational eigenmode as
follows: we take each of the zero-energy floppy modes created by
cutting bonds, and distort it by a smoothly-varying sine-like
amplitude which vanishes at the boundaries of the box, right at the
point where we have artifically cut the bonds. This long-wavelength
distorted mode is a good Ansatz for a low-energy eigenmode: when we
analyze the potential energy associated with this Ansatz-mode, the
underlying floppy mode does not contribute to the energy --- only
the fact that the mode has been made imperfect by the elastic
distortion on the scale $\ell$ contributes at every bond a
distortion energy of order $1/\ell^2$ (nearby displacements differ
from that in the underlying floppy modes by an amount of order the
gradient of the amplitude, so the energy, which involves terms of
order the average compression squared, is of order $1/\ell^2$).
Hence this Ansatz mode will have a frequency $\omega_\ell ={\mathcal
O}(1/\ell)$.

Of course, in a variational calculation, each of these Ansatz modes
will acquire a lower energy upon relaxation, but the number of them
will not change (rigorously speaking this argument yields a lower
bound on the density of states).  In what follows we shall call the
modes obtained from the distortion of floppy modes `anomalous', as
their nature is very different from that of plane waves.  Assuming
that the energy does not shift dramatically, we have created $
N_\ell \simeq \ell^{d-1}$ modes with frequency up to
$\omega_\ell\sim \ell^{-1}$ in a box with of order $V_\ell\sim
\ell^d$ particles. Thus we have
\begin{equation}
\int_0^{\omega_\ell} {\rm d}\omega D(\omega) \simeq
\frac{N_\ell}{V_\ell} \sim \frac{1}{\ell}.
\end{equation}
If we assume that $D(\omega)$ scales as $\omega^q$ for small
frequencies, we get
\begin{equation}
(\omega_\ell)^{q +1} \sim \frac{1}{\ell^{a+1}} \sim \frac{1}{\ell},
\end{equation}
which immediately yields $a=0$: $D(\omega)$ is flat at low
frequencies.

When the packings are compressed so that they have an excess number
of bonds $\Delta Z = Z-Z_c$, the above line of reasoning can also be
followed to obtain the cross-over frequency $\omega^*$ and length
scale $\ell^*$ that separate the range dominated by isostaticity
effects from the usual elastic behavior. Instead of starting from an
isostatic packing, we retrace the above construction for a packing
with given excess contact number $\Delta Z$ per particle. The total
number of excess bonds within the box of Fig.~\ref{jamfig:4}(b) then
scales like $\Delta Z \, \ell^d$. If the number of degrees of
freedom $N_\ell \simeq \ell^{d-1}$ created by cutting bonds at the
boundary of the box is less than the excess number of bonds in the
bulk, no zero-energy modes will be created. This fact was already
noticed by Tkachenko and collaborators in inspiring works
\shortcite{tkachenko1,tkachenko2}.  Hence we expect a cross-over
scale $\ell^*$ when the two terms balance,{\em i.e.}, when
\begin{equation}
N_{\ell^*} \simeq (\ell^*)^{d-1} \simeq \Delta Z\, (\ell^*)^{d} ~~~
\Longrightarrow ~~~ \ell^* \sim 1/ \Delta Z. \label{ellz}
\end{equation}
Likewise, according to the argument above, one expects the crossover
frequency $\omega^*$ in the density of states above jamming will
scale as
\begin{equation}
\omega^* \sim 1/\ell^* \sim \Delta Z. \label{omegaz}
\end{equation}

\begin{figure}[t]
\centering
\includegraphics*[width=.99\textwidth]{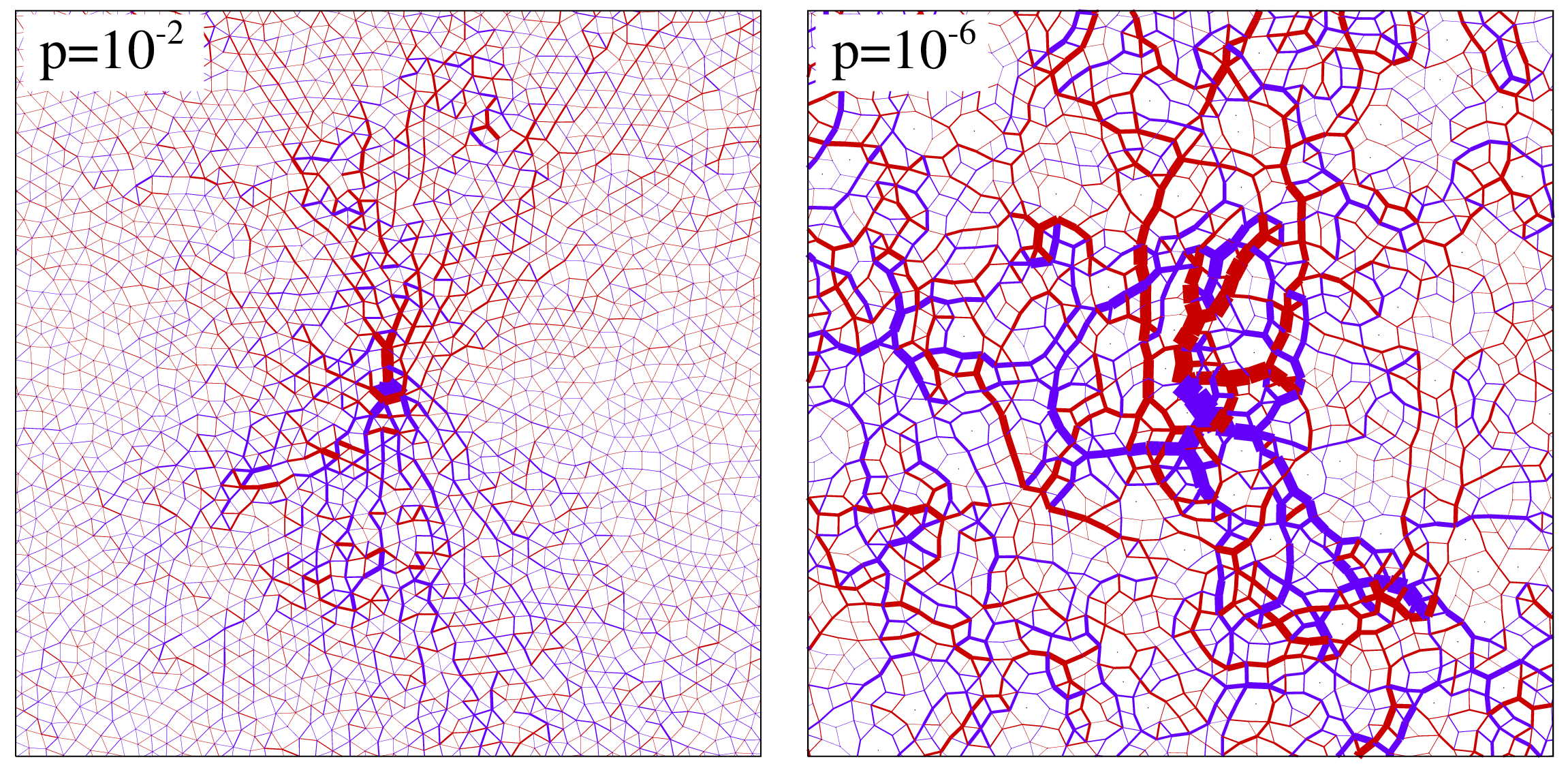}
\caption[]{ Illustration of the fact that the length scale $\ell^*$
increases as the jamming threshold is approached. This figure
illustrates the response to a force loading in the center, with blue
(red) lines indicating an increase (decrease) of the force at a
contact; the thickness of the lines is proportional to the size of
the change in force. The left panel is for a pressure $p=10^{-2}$,
the right one is close to the jamming point, {\em i.e.}, at pressure
$p=10^{-6}$. In the latter case, the fluctuations are larger and
extend over a larger region. A detailed analysis shows that this
range grows proportionally to the length scale $\ell^*$. From
Ellenbroek {\em et al.}
\protect\shortcite{respprl,EllenbroekPRE09}.} \label{jamfig:5}
\end{figure}

The scaling of the cross-over frequency, $\omega^*\sim \Delta Z$,
has been well documented from a detailed analysis of the density of
states, as shown in Fig.~\ref{jamfig:2}.  At finite compression,
$\omega^*$ is the frequency above which there is a plateau in
$D(\omega)$, and below which $D(\omega)$ decreases with decreasing
frequency. Early data obtained as a function of excess density
$\Delta \phi$ are shown in Fig.~\ref{jamfig:3}(b), and are found to
scale as $\omega^*\sim |\Delta \phi|^{0.5}$.  Since $\Delta Z \sim
|\Delta \phi|^{0.5}$, see Fig.~\ref{jamfig:3}(a), this is consistent
with the scaling $\omega^* \sim \Delta Z$. An explicit plot of
$\omega^*$ versus $\Delta Z$ can be found in \shortcite{wyartEPL}.

We stress that in the above argument, it is implicity assumed that the bond strength $k$ remains unchanged. This is true for one-sided harmonic forces, but not for Hertzian interactions which weaken upon approaching the jamming point: $f\sim \delta^{3/2}$ so that $k\sim \delta^{1/2}\sim \sqrt{\Delta \phi}$. Since frequencies scale as $\sqrt{k}$, the above arguments generalize to non-harmonic forces if formulated in terms of scaled frequencies $\tilde{\omega}=\omega/\sqrt{k}$. 

While the cross-over frequency is relatively easy to extract from
the vibrational density of states, the cross-over length $\ell^* $
is more difficult to extract from a mode or from response data.
Physically, $\ell^*$ is the length scale on which the
lowest-frequency anomalous modes probe the microscopic structure of
the solid, and are therefore sensitive to its fluctuations. Thus one
may expect to observe $\ell^*$ in the fluctuations of the linear
response of the solid, rather than in its mean behavior.  Another
difficulty lies in the fact that the floppy modes which form the
basis for the behavior up to scale $\ell^*$ are very disordered so
that a weak, elastic-like distortion of such modes is difficult to
detect. Nevertheless, it has been discovered that the cross-over
length is easily discernable by eye in the response to a point force
or to inflation of a local particle
\shortcite{respprl,EllenbroekPRE09}, as Fig.~\ref{jamfig:5}
illustrates. A detailed analysis of these data has indeed shown that
the response is governed by a crossover scale that grows as
$1/\Delta Z$, in accord with (\ref{ellz}).

\subsection{Microscopic criterion for stability under compression}\label{sectionmatt}

So far our analysis of the density of states has neglected the
effect of pressure on the vibrational spectrum. Mundane observations
such as the buckling of a straw pushed at its tips show that
compression affects vibrational modes, and can even make them
unstable. In general, this effect is important for thin objects, but
is irrelevant at small strain for bulk solids where plane waves
dominate the vibrational spectrum. However compression plays an
important role if floppy modes are present in the solid, as noticed
by Alexander in the context of gels \shortcite{alexander}. As we now
show, this is also true for the anomalous modes introduced above
\shortcite{wyartPRE}.

\begin{figure}[t]
\centering
\includegraphics*[angle=-90,width=.49\textwidth]{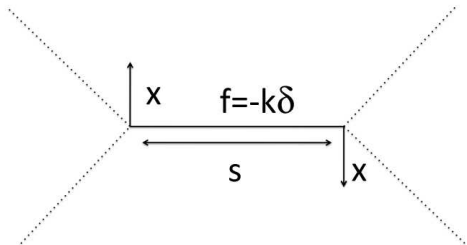}
\caption[]{Consider for concreteness a harmonic contact between two
particles of length $s$, of stiffness $k$, carrying a force
$f=-k\delta$ where $\delta$ is the contact elongation. If particles
moves transversally to the contact by an amplitude $x$ as shown with
vertical arrows, the contact length increases by an amount
proportional to $x^2/s$ following Pythagoras theorem. The work
produced by the contact force is then proportional to $x^2k\delta
/s$, corresponding to a stiffness $k_1\sim k \delta/s\equiv-k e$,
where $e\equiv-\delta/s$ is the contact strain, chosen to be
positive for compressed contacts. } \label{jamfig:5bis}
\end{figure}


The analysis is based on the well-known observation, see
Fig.(\ref{jamfig:5bis}), that if $k$ is the stiffness associated
with the longitudinal relative displacements of two particles in
contact, then for transverse displacements the stiffness is
$k_1\propto -k e$, where $e$ is the contact strain.  For a pure
plane wave propagating in an amorphous solid, the transverse and
longitudinal components of the relative displacements in the
contacts are of the same order. The correction induced on the mode
energy by the presence of a finite contact force or strain is thus,
in relative terms, of order of the ratio of the two stiffnesses:
$-e$. Therefore the effect of compression on plane waves is
negligible at small strain. For anomalous modes the situation is
different, because they are built by deforming floppy modes that
have no longitudinal relative displacements, but only transverse
ones.  Once floppy modes are deformed to generate anomalous modes,
they gain a longitudinal component of order $\Delta Z$ as follows
from the variational argument, whereas the transverse component
remains of order one \shortcite{wyartPRE,wyartthesis}. Relative
corrections in the anomalous mode energies is thus of order $-
e/\Delta Z^2$. When these relative corrections reach a constant of
order one, anomalous modes become unstable, and the system yields.
For particles near jamming, one has $e\propto  \Delta \phi$, leading
to the condition for stability \shortcite{wyartPRE}:
\begin{equation}
\label{maxext} \Delta Z > \Delta \phi^{1/2} ~~~~~ \mbox{for all
subsystem of size}~ L>l^*
\end{equation}
where a numerical pre-factor is omitted. Inequality(\ref{maxext})
extends Maxwell criterion to the case of finite compression and is
also not a local criterion: it must be satisfied on all subsystems
of size larger than $l^*$. On smaller scales, fluctuations of
coordination violating Eq.(\ref{maxext}) are permitted, as stability
can be insured by the boundaries.
Configurations where the bound (\ref{maxext}) is saturated are marginally stable and 
anomalous modes exist down to zero frequency.  Such packings should
exhibit the scaling $\Delta Z \sim \Delta \phi^{1/2}$.
Configurations obtained by decompression in the vicinity of $\phi_c$
appear to lie very close to saturation \shortcite{wyartPRE}.  The
scaling of Eq.(\ref{maxext}) was proposed independently by Head
\shortcite{head} following a mean-field analysis; there, however,
the threshold coordination found was half its correct value.

An argument for why one might expect the system to be marginally stable is provided in section \ref{HS}, where it will be argued that the realization of the bound (\ref{maxext}) critically affects the dynamics.  The criterion (\ref{maxext}) applies to spheres, but not to ellipses where rotational degrees of freedom matter. 
It can be shown that in the later case, compression can have a {\it
stabilizing} effect if the ellipses make in average more contacts
where their surface curvature is low. In that case, even an
infinitesimal pressure can stabilize zero-modes, so that stable
packings can be generated although the Maxwell bound is not
fulfilled, as shown in section \ref{jamsection:3.1}. The same is
true for gels of cross-linked polymers \shortcite{alexander}.

\subsection{Behavior or elastic constants near jamming}\label{sectionelastic}

The fact that a packing of frictionless spheres or discs is a
marginally-connected solid, also has its effect on the elastic
constants. The behavior of the bulk compression modulus $B$ and
shear modulus $G$ always depends on the specific force law, and
hence is not universal: for distributions of bonds which are peaked
around some average non-zero value, both of them are proportional to
the average bond strength $k$, hence in discussing the effects due
to jamming it is useful to divide out this common effect. The
individual bond strength $k_{ij}$ are essentially the second
derivatives of the interaction potential evaluated at each contact
--- for the one-sided harmonic springs these are all the same, but
for the Hertzian $f_{ij}\sim \delta_{ij}^{3/2}$ force law one finds
$k_{ij}\sim \delta_{ij}^{1/2} \sim f^{1/3}_{ij}$, so that the
average bond strength scales with the pressure $p$ as $p^{1/3} $.

\begin{figure}[t]
\centering
\includegraphics*[width=.9\textwidth]{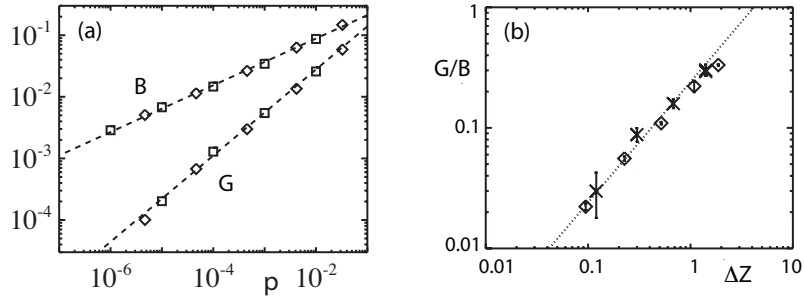}
\caption[]{(a) As the jamming point is approached, in this case by
lowering the pressure, the shear modulus $G$ becomes much smaller
than the bulk compression modulus
$B$~\protect\shortcite{durian95,epitome}. These results are for
packings with Hertzian forces, for which the bond strength of
individual forces scales as $p^{1/3}$. Due to this weakening of the
bonds, the bulk modulus $B$ has an overall $p^{1/3}$ scaling, while
the shear modulus $G$ has an overall $p^{2/3}$ scaling. (b) The
ratio $G/B$ is linear in $\Delta
z$~\protect\shortcite{durian95,epitome}; this scaling is independent
of the force law. Squares refer to data obtained from response to a
local point force, diamonds to data obtained from a global
deformation. After \protect\shortcite{respprl,EllenbroekPRE09}.}
\label{jamfig:6}
\end{figure}

An example of data for the elastic moduli is shown in Fig.~\ref{jamfig:6}, which summarizes data for $B$ and $G$ for packings of two-dimensional discs interacting with Hertzian forces \shortcite{respprl,EllenbroekPRE09}. The bulk modulus $B$ is clearly seen to scale as $p^{1/3}$~\shortcite{epitome}, which as we argued above is the scaling of the average bond strength for this force law. However, $G$ decreases much faster, as $p^{2/3}$~\shortcite{epitome}: packings close to jamming are much easier to distort by shear than by compression. Results in both two and three dimensions for harmonic~\shortcite{durian95,epitome} and Hertzian potentials~\shortcite{epitome} have been determined
.  From these results, it is generally seen that $G/B$ and $\Delta
Z$ scale in the same way with compression.  The data in
Fig.~\ref{jamfig:6} are consistent with this, as for a Hertzian
interaction, $G/B \sim p^{1/3} $ and $\Delta Z \sim |\Delta
\phi|^{0.5} \sim p^{1/3}$.  The result that the shear modulus
generally scales as $G\sim k\, \Delta Z$, can be
understood\shortcite{wyartthesis}: independent of the force law, the
ratio $G/B$ should scale as $G/B\sim \Delta Z$.  Empirical support
for this behavior in grains is reviewed in \shortcite{agnolin}.

It is interesting to note that at first sight the fact that $G/B$
goes to zero at the jamming point as $\Delta Z$, may seem to be the
anomalous behavior; however, from the point of view of rigidity
percolation \shortcite{thorpe95}, $B/k$ behaves anomalously for
packings, as in rigidity percolation both $G/k$ and $B/k$ vanish as
$\Delta Z$, upon approaching the isostatic point so that $G/B$
remains constant at the percolation threshold.
The difference between the jamming transition and generic rigidity percolation is that in the former case configurations must be such that all the contact forces 
must be repulsive.  This imposes a geometric constraint on the
network that is not enforced in standard rigidity percolation
models~\shortcite{wyartthesis}.  We refer to
\shortcite{ellenbroekEPL09} for further details.

To compute static properties such as $G/B$ theoretically, one may
use effective medium theory. This approach has been used to study a
square lattice with randomly placed next-nearest-neighbor
springs~\shortcite{xiaoming} and random isotropic off-lattice spring
networks~\shortcite{GarbocziThor1985,matthieucpap}.   In the
isotropic case (rigidity percolation), $G/B$ is independent of
$\Delta Z$ at the transition, as found in
Ref.~\shortcite{ellenbroekEPL09}, while in the randomly decorated
square lattice $G/B \sim \Delta Z^2$.  Both results differ from the
scaling observed for jammed packings, $G/B \sim \Delta Z$.

An important topic in granular research has been the issue of the
coarse graining scale needed for elastic behavior to emerge in
granular packings \shortcite{goldcoarse,gold08}. In line with the
arguments given above, it has been found that near the jamming
point, one needs to coarse grain over the scale $\ell^*$ to see
elastic response emerge \shortcite{respprl,EllenbroekPRE09}: as one
approaches the jamming point, one has to coarse grain over larger
and larger lengths, diverging as $\ell^*\sim 1/\Delta Z$. That
elasticity emerges in granular media on long enough scales is also
implicitly confirmed by the data of Fig.~\ref{jamfig:6}, where
squares give the values of the elastic constants obtained from local
point response calculations, and diamonds those obtained from global
deformations. The two sets of data agree very well
\shortcite{respprl,EllenbroekPRE09}.

\subsection{Diffusivity, quasilocalized modes and anharmonicity}\label{sectiondiffusion}

\subsubsection{Diffusivity}

The excess low-frequency vibrational modes associated with the
isostatic jamming transition have important consequences for the
nature of energy transport in the system.  Their contribution to
energy transport can be quantified by the energy diffusivity,
defined as follows~\shortcite{allen89,allen93}.  Consider a
wavepacket narrowly peaked at frequency $\omega$ and localized at
$\vec r$. This wavepacket spreads out in time, such that the square
of the width of the wavepacket at time $t$, divided by $t$, is
independent of $t$ at large $t$; this constant defines the
diffusivity, $d(\omega)$.  The diffusivity was calculated using the
Kubo approach within the harmonic approximation for repulsive sphere
packings as a function of compression in \shortcite{Xu09,Vitelli09}.
At sufficiently low frequencies, the modes should be plane-waves and
the diffusivity should follow Rayleigh scattering behavior so that
$d(\omega) \sim \omega^{-4}$.   Refs.~\shortcite{Xu09,Vitelli09}
show that the diffusivity is nearly flat for frequencies from
$\omega^*$ up to the Debye frequency, where it drops to zero.  Thus,
there is a crossover from weakly-scattered plane waves to
strong-scattering, diffusive behavior at $\omega^*$; this can be
understood from the heterogeneous nature of the anomalous modes
\shortcite{wyartthesis,Vitelli09}, or by assuming the continuity of
the length scale characterizing plane waves and anomalous modes at
the cross-over frequency, in conjunction with the scaling laws for
the shear and bulk moduli discussed in section~\ref{sectionelastic}
\shortcite{Vitelli09,Xu09}.

\begin{figure}[t]
\centering
\includegraphics*[width=.42\textwidth]{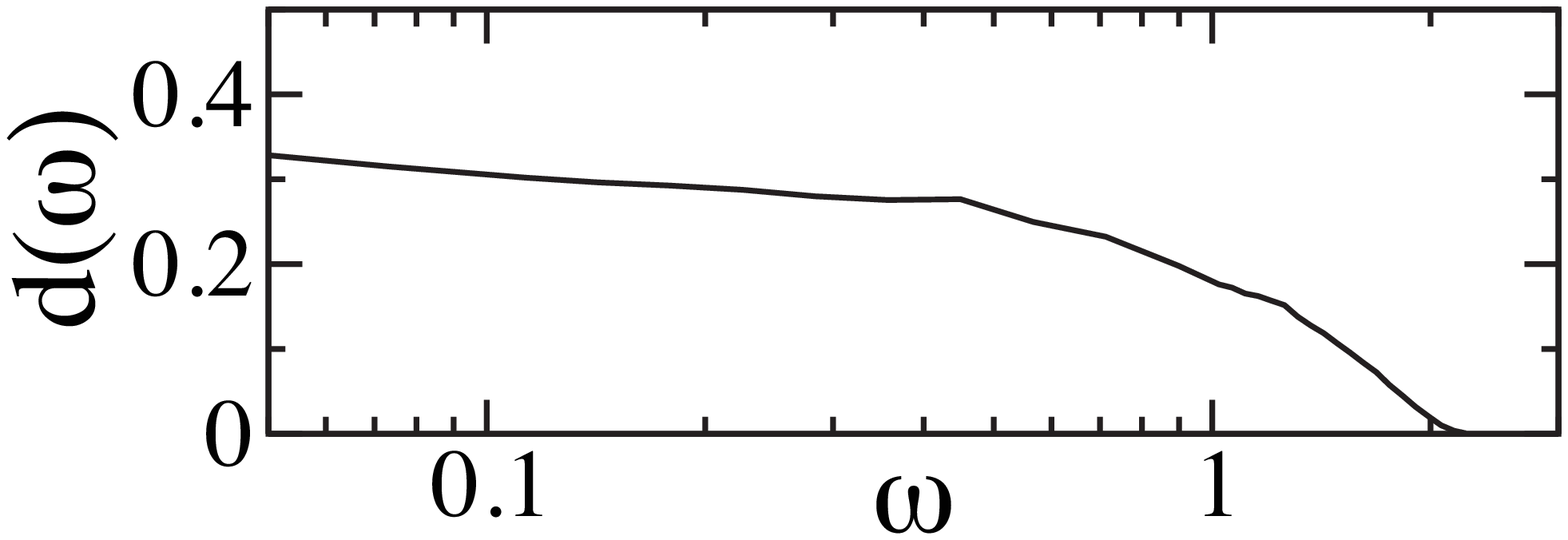}
\includegraphics*[width=.5\textwidth]{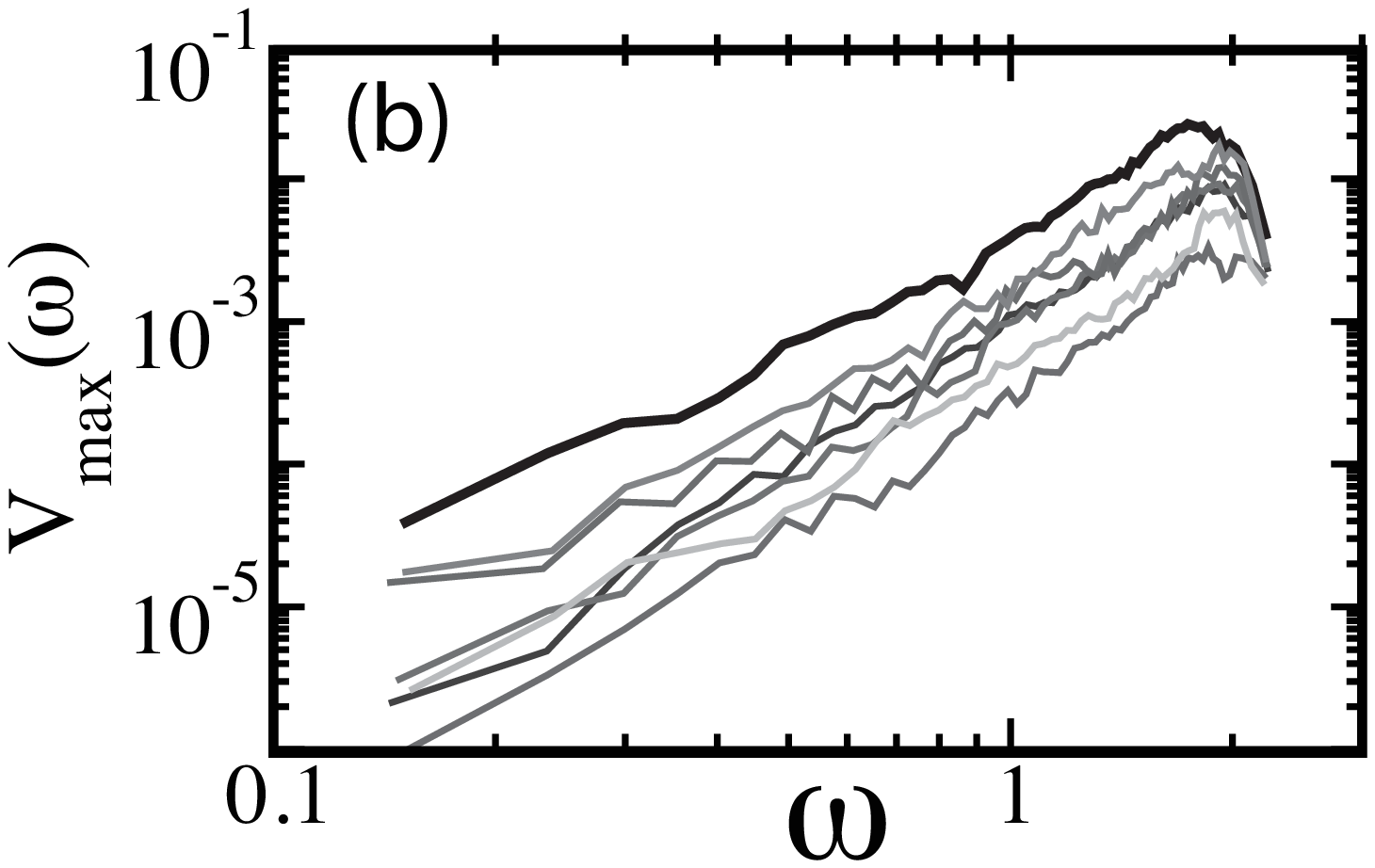}
\caption[]{(a) Diffusivity, $d(\omega)$, for a system of spheres
interacting via harmonic repulsions very close to the jamming
transition. $d(\omega)$ is essentially flat down to zero frequency.
From Xu {\em et al.} \protect\shortcite{Xu09}. (b) Energy barrier
$V_{\rm max} (\omega)$ along each mode, before falling into another
basin of attraction for a system at $\Delta \phi=0.1$.  The energy
barriers along each mode direction provide an upper bound on the
energy barriers of the system in a given frequency range, since it
is possible that lower energy barriers could be encountered in
directions in phase space corresponding to linear combinations of
the modes. From \protect\shortcite{xu09b}.
 }
\label{jamfig:7}
\end{figure}

One consequence of the link between the transport crossover and the
excess modes~\shortcite{Xu09,Vitelli09} is that the diffusivity
plateau extends all the way down to the lowest frequencies studied
as the jamming transition is approached and $\omega^* \rightarrow
0$, as shown in Fig.~\ref{jamfig:7}(a).  This suggests that the
origin of the diffusivity plateau can be traced to properties of the
jamming transition.


Dynamical vibrational properties such as the energy diffusivity can
be computed theoretically using dynamical effective medium theory, a
re-summation scheme of a perturbation expansion in the disorder also
known as the Coherent Potential Approximation (CPA)
~\shortcite{Soven}. This approach has been used on the square
lattice with randomly placed next-nearest-neighbor
springs~\shortcite{xiaoming} and isotropic disordered spring
networks near the isostatic limit~\shortcite{matthieucpap}.  Both
calculations yield a crossover to a flat density of states above
$\omega^\star \sim \Delta Z$ in good agreement with results for
jammed packings. In the isotropic case, Rayleigh scattering is found
for $\omega < \omega^\star$ with a scattering length $\ell_s \sim
\Delta Z^3/\omega^4$.  This calculation predicts a violation of the
Ioffe-Regel criterion (according to which the cross-over to strong
scattering occurs when the scattering length becomes of order of the
wavelength) at the crossover frequency $\omega^\star$ since the
scattering length $\ell_s \sim 1/\Delta Z$ is much larger than the
wavelength $\lambda \sim 1/\sqrt{\Delta Z}$ there.  Above
$\omega^\star$, the anomalous modes are characterized by a
displacement-displacement correlation length that scales as
$1/\sqrt{\omega}$, which scales as $1/\sqrt{\Delta z}$ at
$\omega^\star$, and a frequency-dependent speed of sound, $c(\omega)
\sim \sqrt{\omega}$. This leads to a plateau in the diffusivity
consistent with the results of Fig.(\ref{jamfig:7}).
This analysis predicts a 
drop 
in the speed of sound near $\omega^*$, a subtle effect recently
observed numerically at the Boson peak frequency in model glasses
\shortcite{mossa}. In the case where the isostatic state is a square
lattice~\shortcite{xiaoming}, the Ioffe-Regel criterion is satisfied
at the crossover frequency $\omega^*$ and $\ell_s$ is of order the
lattice spacing.  Numerical results for jammed packings are
consistent with $\ell_s \sim 1/\sqrt{\Delta z}$ at the crossover.

\subsubsection{Quasilocalized modes}

The nature of the excess low frequency modes shows a systematic
variation with frequency.  As the frequency is lowered towards
$\omega^*$, numerical studies of three-dimensional sphere packings
show that the modes become progressively more heterogeneous with a
lower-than-average mode coordination number, and develop high
displacement amplitudes in small regions of space~\shortcite{Xu09}.
These quasi-localized modes shift downwards with decreasing
compression.   These low-frequency, quasi-localized modes have also
been observed in two-dimensional packings of soft
disks~\shortcite{zeravcic09}, but here the quasi-localized modes are
mixed in with plane-wave-like modes over the same low-frequency
range.

In three dimensions, the frequency of the quasi-localized modes
decreases with compression, but the degree of quasi-localization
does not appear to depend on compression.  This suggests that as the
system is decompressed towards the unjamming transition,
quasi-localized modes always exist but shift to lower and lower
frequency.  Quasi-localized modes have been observed in other
disordered models~\shortcite{biswas,fabian,taraskin,SchoberRuocco}
and in experiments on glassy
polymers~\shortcite{buchenau92,buchenau96,vainer06}.    In the
soft-potential model, for example, such modes have been shown to
lead to strong scattering of vibrations at higher frequencies, so
that their frequency corresponds to the Ioffe-Regel crossover
frequency in the transport properties~\shortcite{schober}, while in
model silica, these modes have been identified near the boson peak
frequency~\shortcite{taraskin}.

Numerical studies on hard spheres~\shortcite{brito07b,brito3} have
shown that irreversible particle rearrangements at nonzero
temperature occur along the low-frequency modes,
 observed to be extended in the systems of limited  size studied, implying that the lowest activation barriers are in those directions of phase space (Sec.~\ref{HS}).  This view has been explored in model glasses~\shortcite{harrowell}, which have also shown that the structural rearrangements occurring at finite temperatures were along low-frequency normal modes, which in that case were quasi-localized.

Fig.~\ref{jamfig:7}(b) shows that for sphere packings, the
low-frequency quasi-localized modes have the lowest energy barriers
and that the energy barriers decreases systematically with frequency
and participation ratio.  This result,
together with previous observations
\shortcite{brito07b,brito3,harrowell},
 should have important consequences for the dynamics and instabilities that occur as the temperature is raised from zero, the system is compressed, or a shear stress is applied.
One would expect these modes to shift downwards to zero frequency
under the last two circumstances~\shortcite{maloney} and to give
rise to highly heterogeneous deformations in the sample.  These
could be the underlying mechanism for the localized deformations
postulated in the picture of shear transformation
zones~\shortcite{langer98,langer08}.

As the system is decompressed towards the unjamming transition, the
barriers between nearby configurations shrink to zero as the amount
of inter-particle overlap decreases to zero.  Thus, in this limit
the modes become progressively more anharmonic.  This implies that
smaller excitation energies or temperatures should suffice to force
the system into new energy minima.  Thus, the range of temperatures
and stresses over which harmonic theory applies should shrink to
zero and anharmonic effects should dominate as one approaches the
transition.
This effect has been computed in networks of springs below or above
the isostatic threshold \shortcite{wyart08}, where non-linearities
dominate the response to shear when the strain $\gamma$ is larger
than some $\gamma^*$, with $\gamma^*\sim|\Delta z|$.

\subsection{Structural features of the jamming transition}\label{sectionstructural}

At point J, the pair distribution function has many singular
properties \shortcite{epitome,silbertPRE06}.  In particular, it has
been shown that the first peak in $g(r)$ is a $\delta$-function at
the nearest neighbor distance, $\sigma$.  The area under this
$\delta$-function is just the isostatic coordination number: $Z = 6$
in three dimensions.  On the high side of this $\delta$-function,
$g(r)$ has a power-law decay: $g(r) \propto (r-\sigma)^{-0.5}$. It
has been proposed that this power-law is the vestige of the marginal
stability of the configurations visited before reaching $\phi_c$
\shortcite{wyartthesis}.
 Also, at the jamming transition, the second peak of $g(r)$ is split into what appear to be two singular sub-peaks Ð one at $r= \sqrt {3}\sigma$ and the other at $r=2\sigma$.  There is a divergent slope just below each subpeak and a step-function cutoff on its high side\shortcite{silbertPRE06}.

 The pair correlation function near the zero-temperature jamming transition has been studied in two experiments.  In experiments by Abate and Durian~\shortcite{abatePRE06,keysNP07}, a layer of ball bearings were placed on a wire mesh and excited by the upflow of gas through the mesh.  The turbulent flow of the gas leads to interactions between the balls and to random motion of the balls.  Thus, the effective interaction potential between the balls can be characterized by a hard core determined by the ball diameter and a longer-ranged repulsion due to the gas.  In such systems, the kinetic energy of the balls decreases monotonically with increasing density of the balls, such that it reaches zero when the balls are packed to the density of the jamming transition, $\phi_c\simeq 0.84$, measured in terms of the hard-core diameter.  The structure of the system, as characterized by the pair correlation function, $g(r)$, was studied along this trajectory to the zero-temperature jamming transition of the hard cores (See the chapter in this book by Dauchot, Durian and van Hecke for a description of the dynamical properties of this system).  On approaching $\phi_c$, the structure shows an increase of the height of the first peak, $g_1$.  This is consistent with numerical results on soft spheres, which find a divergence in $g_1$ at the zero-temperature jamming transition.  There is also a local maximum in $g_1$ at $\phi \approx 0.74$, above which no changes in the coordination and geometrical features of Voronoi cells are observed.  This secondary maximum at nonzero kinetic energy due to airflow-induced interactions, may correspond to the finite-temperature vestige~\shortcite{arjun} of the divergence in the height of the first peak of $g(r)$ at the zero-temperature transition discussed below.

 \begin{figure}[t]
\centering
\includegraphics*[width=.8\textwidth]{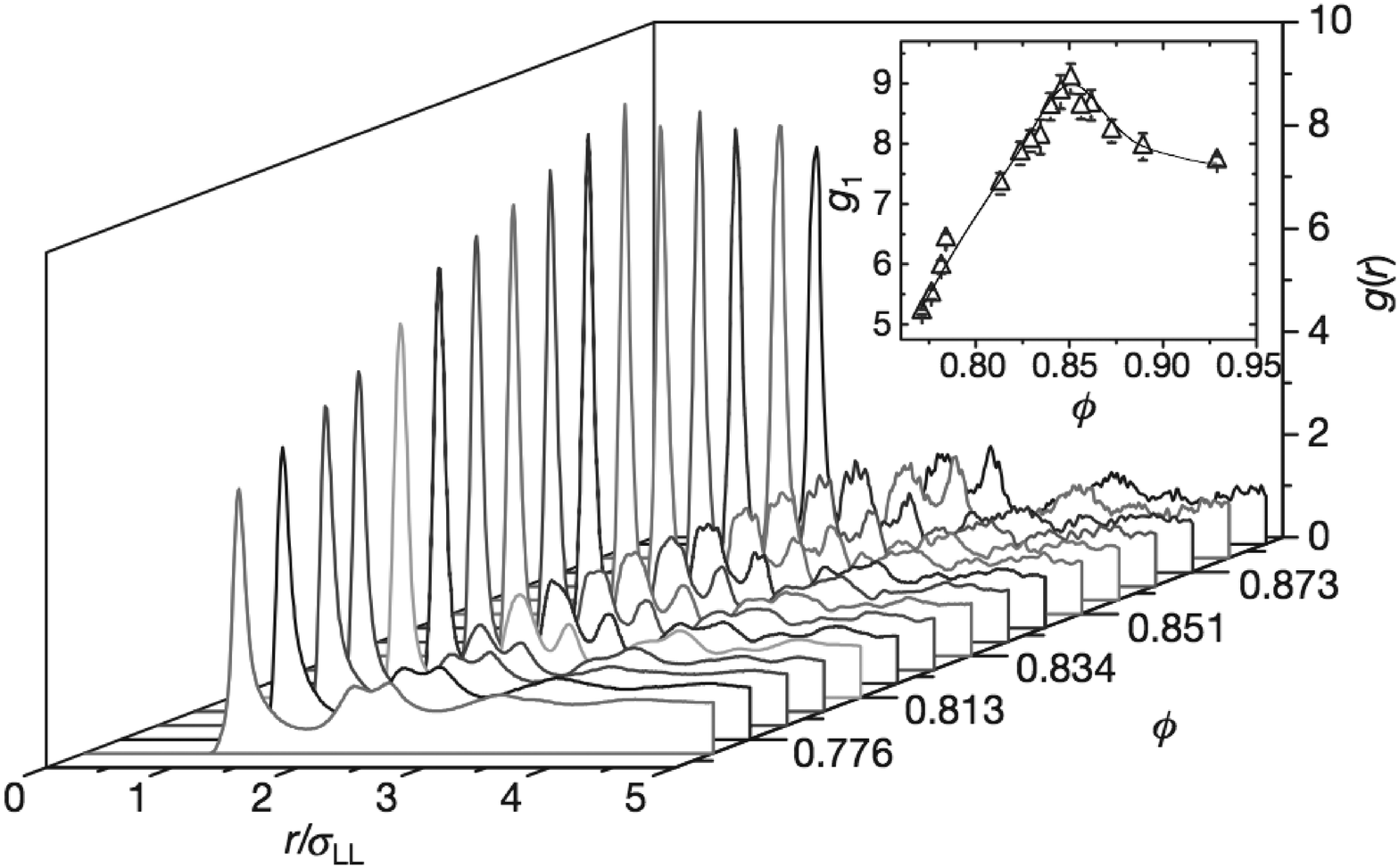}
\caption[]{Pair-correlation function $g(r)$ for the large particles
at different packing fractions $\phi$ for a bidisperse system of
colloids in two dimensions.  The inset shows $g_1$, the height of
the first peak of $g(r)$, versus packing fraction.  The experimental
data shows that $g_1$ is non-monotonic with a maximum at a packing
fraction slightly above $\phi_c$, the value at the $T=0$ jamming
transition.  The peak is a vestige of that $T=0$ transition where
the first peak of $g(r)$ is a $\delta$-function.  Figure reproduced
from  \protect\shortcite{arjun}.} \label{jamfig:12}
\end{figure}

Experiments on a two-dimensional bidisperse system of NIPA particles by Zhang, et al.~\shortcite{arjun}, measured the pair correlation function as a function of density at essentially fixed, nonzero temperature.  The system exhibits a maximum in the height of the first peak of the pair correlation, $g_1$ --- see Fig.~\ref{jamfig:12}.  This maximum is interpreted as a thermal vestige of the zero-temperature jamming transition
.  Corroborating simulations show~\shortcite{arjun} that the
divergence in $g_1 \sim 1/(\phi-\phi_c)$ for soft repulsive spheres
at the zero-temperature jamming transition is softened to a maximum
at nonzero temperature.  The maximum in $g_1$ decreases in height
and shifts to higher density with increasing temperature.  The
simulations also show that the maximum in $g_1$ is a structural
feature that is apparent as a function of increasing density at
fixed temperature, but not as a function of temperature at fixed
density or pressure.  Recent simulations on the same model by
Jacquin, et al.~\shortcite{jacquin} show that the maximum in $g_1$
appears even in systems at higher temperature or lower pressure that
are in thermal equilibrium and can be understood from liquid state
theory.

Note that the divergence in $g_1$ corresponds to a vanishing of the overlap distance between neighboring particles.  Thus, tracking $g_1$ is equivalent to tracking the overlap distance.  The evolution of the overlap distance with temperature is not correlated with the existence of a dynamical glass transition.  

\subsection{Singular length scales at Point J}

Several different singular length scales have been identified near the jamming transition.  The cutting length, $\ell^*$, discussed in Sec. 1.2.3, has been observed by studying the fluctuations in the change of the force on a contact as a function of the distance of the contact from a small local perturbation (see Fig. 1.5).  This length scale diverges as $(\phi-\phi_c)^{-1/2}$, independent of the potential as long as it is repulsive and finite in range.  In addition, this length scale is expected to correspond to the wavelength of longitudinal, weakly-scattered plane waves just below the crossover frequency $\omega^*$~\shortcite{silbertPRL05}, although it has not been seen numerically, 
due to the small system sizes that can be studied.

A second diverging length scale has been identified from the
wavelength of transverse, weakly-scattered plane-waves just below
the crossover frequency
$\omega^*$~\shortcite{silbertPRL05,Vitelli09}. This length scale
diverges as $(\phi-\phi_c)^{-1/4}$.
Finally, this length scale emerges within the effective medium
approximation as the decay length for spatial correlations in modes
at the frequency $\omega \rightarrow
\omega^\star+$~\shortcite{matthieucpap}.

As noted in Sec. 1.2.5, there is a vanishing length scale
corresponding to the overlap distance $\delta$ between neighboring
particles.  This distance, which measures the lefthand width of the
first peak of the pair-correlation function, vanishes as
$(\phi-\phi_c)^{1}$~\shortcite{epitome,silbertPRE06}.

Finally, there are other diverging length scales whose origins are
not understood.  The shift of the position of the jamming
transition, $\phi_c$, with the linear size of the system, $L$,
yields a length scale that apparently diverges as
$|\phi-\phi_c|^{-0.7}$ in both 2 and 3
dimensions~\shortcite{epitome}. This scaling shows up in simulations
in which a hard disk is pushed through a packing below $\phi_c$; the
transverse distance over which the packing adjusts as the disk
passes by diverges with a power-law of 0.7~\shortcite{drocco}.
Finally, this power-law has been observed for correlations of the
transverse velocity on athermal, slowly-sheared sphere packings near
the jamming transition~\shortcite{olsson}.  However, this length
scale is only observed for certain models of the
dynamics~\shortcite{remmers}.

\section{Extensions of the results for frictionless spheres 
} \label{jamsection:3.0}

\subsection{Anisotropic particles: ellipsoids} \label{jamsection:3.1}

The counting of degrees of freedom that underlies the above analysis
of the soft modes near the jamming point, relies on having perfect
spherical and frictionless particles. We briefly review here what
happens to the jamming scenario  if the particles are frictionless
but non-spherical.

The model for which this question has been studied in detail is that
of two-dimensional frictionless ellipses in two dimensions
\shortcite{mailman} or ellipsoids with one axis of symmetry in three
dimensions \shortcite{zeravcic09}. In the latter case, each
ellipsoid is essentially characterized by 5 nontrivial degrees of
freedom, so a naive counting of the degrees of freedom shows that
the isotropic values for such ellipsoids is $Z^{\rm ellipsoid}_{\rm
iso} = 2\cdot 5 = 10$.

Clearly, while the isotropic values jumps depending on which degrees
of freedom are, and which ones are not, included in the counting,
one would expect the physical behavior of such packings to evolve
continuously, when the a-sphericity of the particles is turned on
--- what is happening?

\begin{figure}[t]
\centering
\includegraphics*[width=.98\textwidth]{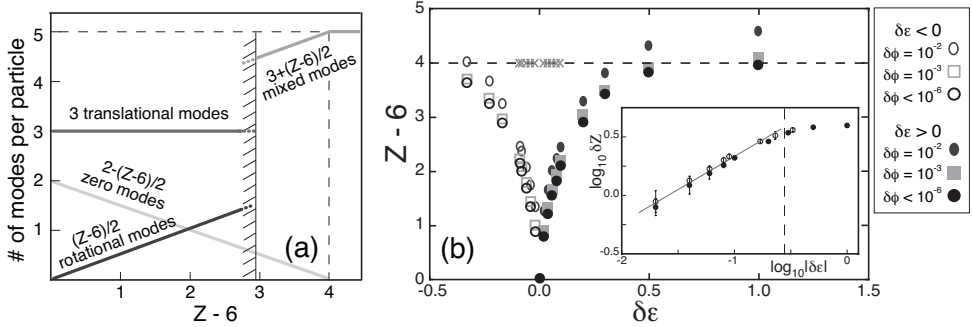}
\caption[]{ (a) Illustration of how the number of different modes
per ellipsoid (excluding rattlers) in the packings at jamming varies
as a function of the average contact number $Z$. As the ellipticity
is increased, the average contact number increases. For $Z< \approx
9$, there are two well-defined bands: a rotational band with
$(Z-6)/2$ modes per particle and a translational band with three
modes per interacting particle as is the case for spheres (see
Fig.~\ref{jamfig:9} for details). Upon increasing $Z$, the number of
zero modes decreases as zero-modes are converted into
finite-frequency rotational modes. Above $Z\approx 9$, there is only
one band. (b) $Z$ as a function of the ellipticity $\delta
\varepsilon \equiv \varepsilon -1 $ and distance from jamming
$\delta \phi$ for $216$-particle packings. The sharp decrease around
the spherical case $\delta \varepsilon=0$ is consistent with earlier
results \protect\shortcite{donev04,sacanna,wouterse} for hard
ellipsoids and spherically capped rods. The log-log plot of $\delta
Z$ versus $|\delta \varepsilon |$ in the inset shows that the rise
of $Z$ at jamming is consistent with a $\delta Z \sim \sqrt{ |
\delta \varepsilon |}$ scaling. The crosses in the main plot for
small values of $\delta \varepsilon$  show that twice the measured
number of zero-frequency eigenmodes per particle plus $Z-6$ add up
precisely to 4, in accord with panel (a). From Zeravcic {\em et al.}
\protect\shortcite{zeravcic09}. } \label{jamfig:8}
\end{figure}

Let us take the $c$ axis of our ellipsoids along their symmetry
axis; the other axes are then $a$ and $b=a$. We define the
ellipticity as the aspect ratio $\varepsilon=c/a$, so prolate
ellipsoids (like cigars)  correspond to $\varepsilon >1$ while
oblate ones (like M\&M's \shortcite{donev04}) to $\varepsilon <1$.
Fig.~\ref{jamfig:8}(b) recovers the finding of
\shortcite{donev04,donev3,sacanna,wouterse} that as the ellipticity
is tuned away from the spherical case $\varepsilon=1$,  the average
contact number indeed increases continuously from the spherical
value $Z_{\rm iso}=6$, as expected.

How to resolve this apparent paradox between the jump in the
counting and the continuity of the physical system? How come one can
reach values of the average contact number below the isostatic value
$Z^{\rm ellipsoid}_{\rm iso}=10$? To answer these questions, it is
good to realize that when we reached the conclusion that the
isostatic value $Z_{\rm iso}^{\rm sphere}=2d=6$, we left out the
rotational degrees of freedom of each sphere, as these are trivial
zero-frequency modes (three per particle) for a system of
frictionless spheres. In order to understand the continuity, it is
better to include these rotational degrees of freedom from the
start, even for the spherical case, and see how they evolve when the
spheres are deformed into ellipsoids. Since in the work
\shortcite{zeravcic09} that we will summarize below, the ellipsoids
still have one symmetry axis, there are only two nontrivial angular
degrees of freedom associated with the anisotropy of these
particles. The relevant isostatic value for these ellipsoids of
revolutation is therefore $Z_{\rm iso}^{\rm ellipsoid}=2\cdot 5=10$,
as already asserted above.

When analyzed more carefully, the counting arguments summarized in
section \ref{sectionisostatic} imply that whenever $Z < Z_{\rm
iso}$, there are $N_c(Z_{\rm iso}-Z)/2$ directions in the phase
space of the relevant coordinates of the particles along which the
packing coordinates can be changed without changing the interaction
energy of the particles. This means that each stable configuration
must have precisely this number of zero-frequency modes. That this
is precisely what happens in the ellipsoid packings, is illustrated
by the grey crosses in Fig.~\ref{jamfig:8}(b); numerically it is
found that in the calculation of the density of states of ellipsoid
packings one finds precisely the right number of zero modes as
expected according to this argument.

Fig.~\ref{jamfig:8}(a) illustrates how, as $Z$ increases with 
$\delta \varepsilon = \varepsilon-1$ of the ellipsoids, the number of zero modes per particle with contacts goes down. We thus have more and more nonzero modes as $Z$ increases. What is their nature? Fig.~\ref{jamfig:8}(a) shows that for 
$|\delta \varepsilon| <  ~0.17$, the density of states exhibits two
clearly separated bands \shortcite{zeravcic09}. The modes in the
upper band of the density of states turn out to be predominantly
translational --- they are very similar to those found in the
sperical packings, and turn out to be characterized by an onset
frequency $\omega^*$ that scales as $\Delta Z = Z-6 \sim |\delta
\varepsilon|^{1/2}$: even though contacts at jamming are created by
deforming the particles and not by compression, the mechanism that
determines the onset of anomalous translational modes at a frequency
$\omega^*$ is the same! Furthermore, the lower frequency band that
exisits for small ellipticity consists of predominantly rotational
modes; it is characterized by an upper frequency that scales
linearly in $| \delta \varepsilon|$ \shortcite{zeravcic09}. For
excentricities larger than about 0.17, the two bands merge into one
mixed band, as Fig.~\ref{jamfig:8}(b) illustrates.

\begin{figure}[t]
\centering
\includegraphics*[width=.9\textwidth]{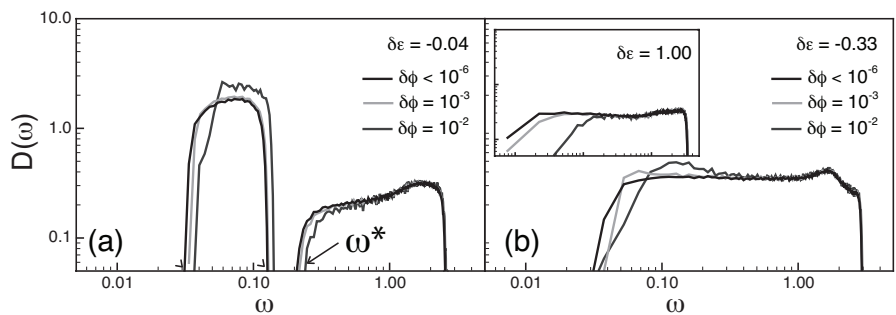}
\caption[]{ (a) The density of states for slightly oblate
ellipsoids, $\delta \varepsilon=- 0.04$, for packings close to
jamming and at two compressions. The existence of two bands
separated by a gap as well as a gap at zero frequency is clearly
visible.  (b) For larger {ellipticities} the two bands merge, as is
illustrated here for $\delta \varepsilon= -0.33$.  The inset shows
the data for $\delta \varepsilon=1$ where $Z\approx Z_{c}=10$ at
jamming (see panel a). In accord with this, the gap near zero
frequency increases with increasing compression (and therefore
increasing Z). From Zeravcic {\em et al.}
\protect\shortcite{zeravcic09}. } \label{jamfig:9}
\end{figure}

The scenario that emerges for the number of modes in the various
bands is summarized in Fig.~\ref{jamfig:9}(a) --- the upshot is that
the jamming scenario as formulated for frictionless spheres, as
exemplified by the sharp onset of anomalous modes, withstands the
change in shape of the particles in the system.

\subsection{Packings of frictional spheres}\label{sectionfriction}

Another relevant extension is to include static friction in the
force law. The usual friction law is to take the Mohr-Coulomb law
\shortcite{johnson}, in which the particles can experience a
tangential force $f_t$ which obeys $f_t \leq \mu f_n$, where $f_n$
is the normal force at the contact considered, and $\mu$ the Coulomb
constant. In the Hertz force-law model, the tangential elastic force
builds up continuously, in a well-described way, when a contact gets
loaded with a tangential force \shortcite{johnson}. This buildup
continues up to $\mu f_n$, beyond that $f_t$ remains constant and
fixed at the value $\mu f_n$. In the presence of static friction,
therefore, the properties of a packing are very history dependent,
as the tangential forces are not only dependent on the
configuration, but also on how the configuration was made --- see
e.g. \shortcite{silbertPRE02jamming,makse05,ellakwave,ellakjam}.

If we return to spherical particles, but include the Mohr-Coulomb
friction law, again the counting of the number of degrees of freedom
is different: the coordinates of the centers of the particles are
still the degrees of freedom with which one can satisfy the ``just
touching conditions'' (\ref{z<2d}) upon approaching the jamming
point. On the other hand, the counting of the degrees of freedom in
the force and torque balance  conditions is different --- e.g., in
two dimensions the frictional forces add one additional  tangential
frictional force component for each contact, but torque balance only
adds one extra constraint per particle. A  straightforward
generalization  (see e.g. \shortcite{kostya}) shows that in general
with frictional forces stable packings can be obtained provided that
$d+1 \leq Z \leq 2d$.  Unlike for the frictionless case, where this
argument dictates $Z$ completely in the limit that the jamming point
is approached, the inclusion of a static frictional force allows a
range of possible values of $Z$. In simulations of frictional
systems, it is indeed indeed found that  packings at the jamming
threshold with different values of $Z$ and density $\phi$ can be
made, by varying in particular the quench rate; moreover, there is a
clear trend for the average contact number $Z$ to approach the lower
lower limit $d+1$ --- the isostatic value in the presence of
friction --- in the limit $\mu \to \infty$ \shortcite{kostya,song}.
In line with this, one finds that if one studies the linear
vibration-rotation  modes of these packings, the density of states
only develops a plateau in this large friction limit
\shortcite{ellakjam}.

 \begin{figure}[t]
\centering
\includegraphics*[width=.32\textwidth]{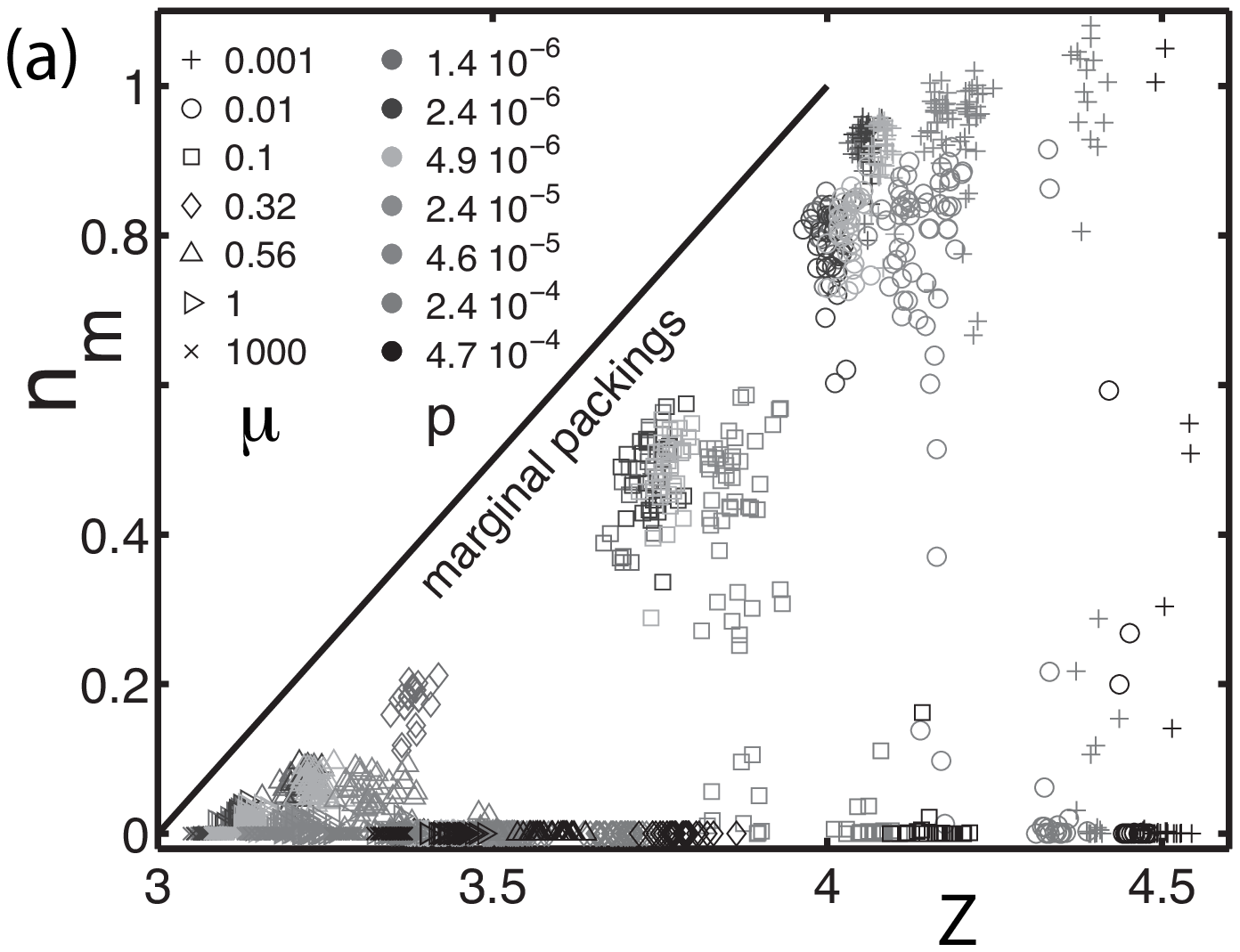}
\includegraphics*[width=.32\textwidth]{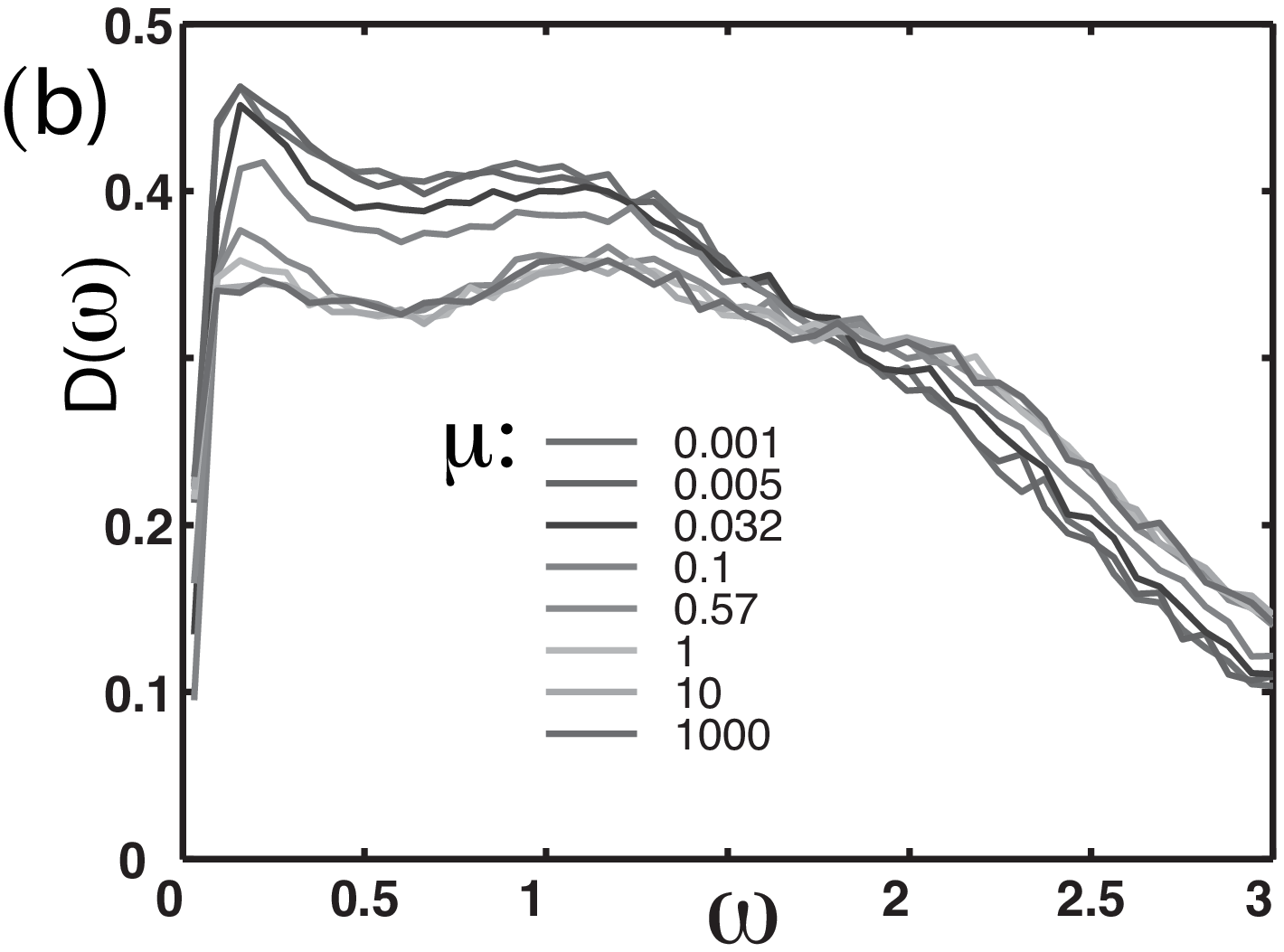}
\includegraphics*[width=.32\textwidth]{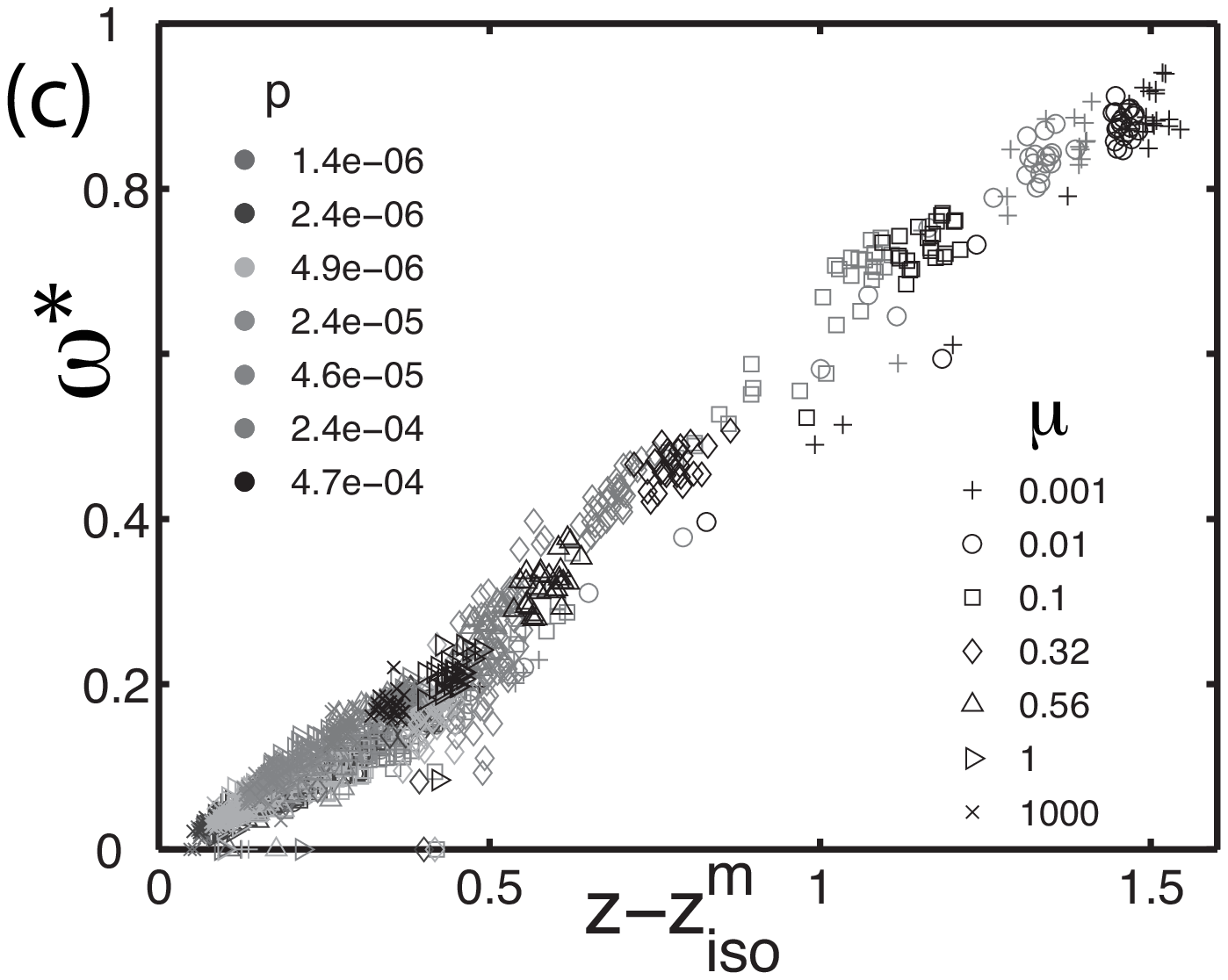}
\caption[]{Data from Henkes {\em et al.}
\protect\shortcite{henkes09} on frictional packings. (a) In gently
prepared frictional packings, there a finite fraction of the
contacts is found to lie at the mobilization threshold (are ``fully
mobilized''), i.e., has $f_t = \mu f_n$. In this diagram a large
number of packing configurations are plotted in a diagram of $Z$ and
$n_m$, the average number of fully mobilized contacts per particle.
For lower pressure, the dots in this diagram representing the
packings approach the generalized isostaticity line, defined by
(\ref{gen_iso}), where packings are marginal. (b) Density of states
for the lowest pressure data for a large range of values of the
friction $\mu$. In all cases the density of states is strongly
enhanced at low frequencies. The frequency plotted here is actually
the scaled frequency $\tilde{\omega}$ mentioned in section
\ref{section1.2.3}. (c) The cross-over frequency $\omega^*$ scales
as $\Delta Z=Z-Z^m_{\rm iso}$. } \label{jamfig:11}
\end{figure}

However, it was later discovered that the situation is more subtle.
If packings are prepared gently, then one finds  that a finite
fraction of the contacts in the simulations are at the mobility or
Coulomb threshold, {\em i.e.}, have $f_t=\mu f_n$
\shortcite{kostya,henkes09}. If such a contact is pushed in the
direction of increasing friction force, they slip! In other words, a
finite fraction of the tangential forces is {\em not} arbitrary, but
constrained to be at the Coulomb threshold. This of course changes
the counting of the number of degrees of freedom. Indeed, if one
introduces $n_m$, the number of contacts per particle which is found
at the mobility threshold, then one may introduce a generalized
criterion \shortcite{kostya,henkes09}
\begin{equation}
Z \geq (d+1) + \frac{2 n_{m}}{d} \equiv Z_{\rm iso}^{m}.
\label{gen_iso}
 \end{equation}
for the stability of packings. In this scenario, packings which are
close to the line $Z = Z_{\rm iso}^m = d+1+2n_m/d$ are again
marginal, {\em i.e.}, develop many low-lying excitations because the
number of contacts that slip under tangential motion is just large
enough to make almost zero-energy deformations possible. The idea
that gently prepared packings might end up being marginal packings
in this way, traces back to an earlier suggestion by Bouchaud
\shortcite{Bouchaud_leshouches}.

With this generalization, packings with static friction that are
gently prepared, do again follow in essence the scenario described
above. As illustrated in Fig.~\ref{jamfig:11}(a), as the pressure on
packings is reduced, the points representing these packings in a
$Z-n_m$ diagram approach the generalized isostaticity line
\shortcite{henkes09}. Moreover, if the contacts at the mobility
threshold are allowed  to slip in analyzing the rotation-vibration
dynamics, the density of states of packings close to the jamming
threshold is strongly enhanced at low frequencies for {\em all}
values of the friction coefficient $\mu$ --- see
Fig.~\ref{jamfig:11}(b). Moreover, as the pressure is increased to
tune the packings away from the generalized isostaticity line, the
crossover frequency $\omega^*$ in the density of states scales
linearly with $\Delta Z=Z-Z^m_{\rm iso}$, {\em i.e.}, linearly with
the distance from the generalized isostaticity line. This is the
first confirmation of the robustness of the concepts reviewed above
to frictional particles.

There are still several unresolved subtleties associated with the
somewhat singular nature of the Coulomb threshold condition, but we
refer for a discussion of these to \shortcite{henkes09}. Another
interesting recent development \shortcite{henkes2010} is that
precursors to avalanches in simulations of tilting of piles of
frictional spheres are found in the correlation properties of balls
with low contact number.

\subsection{Finite shear forces: nontrivial rheology in the sheared bubble model} \label{jamsection:3.2}

So far we have only discussed static packings and their static and
dynamic linear response. An important development of the last two
years is that simulations of bubble models under finite shear stress
or at a constant shear rate are probing the jamming phase diagram
along the stress direction. The bubble model
\shortcite{durian95,durian97} is essentially a soft-sphere model,
enriched with viscous friction, {\em i.e.}, with dynamical terms
proportional to the difference between the velocity of a bubble and
that of each one it makes contact with. A particular attractive
feature of such bubble model simulations with dynamic friction is
the fact that the relevant reference frame for the static limit is
the frictionless sphere or disc model, about which so much is known.

\begin{figure}[t]
\centering
\includegraphics*[width=.5\textwidth]{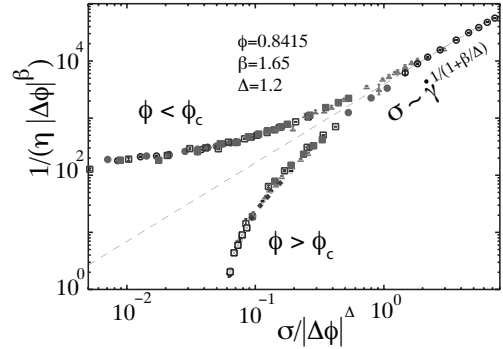}
\caption[]{Scaling plot of Olsson and Teitel
\protect\shortcite{olsson} for simulations of the bubble model under
shear. Plotted is a the viscosity $\eta= \sigma/\dot{\gamma}$,
rescaled by the density, as a function of the applied shear stress
$\sigma$, with $\dot{\gamma}$ the shear rate. The scaling collapse
illustrates that close to the jamming point, nontrivial rheology
occurs. Close to the jamming point, we have $\sigma\sim
\dot{\gamma}^{1/(1+\beta/\Delta)} \simeq \dot{\gamma}^{0.4}$. }
\label{jamfig:10}
\end{figure}

Olsson and Teitel \shortcite{olsson} were the first to demonstrate
that the rheological data for shear flows in such systems show very
good data collapse close to the jamming point of the frictionless
case. Fig.~\ref{jamfig:10} shows a result from their simulations. In
this particular plot, the effective viscosity $\eta=\sigma /
\dot{\gamma}$ is plotted as a function of the shear stress $\sigma $
imposed on the system; here $\dot{\gamma}$ is the shear rate. By
scaling both quantities properly with the distance $\Delta \phi$
from the jamming density, there is very good data collapse both
above and below the critical density. Moreover, at the jamming
density, and close to the jamming density for large enough shear
rates $\dot{\gamma}$, the data is consistent with a scaling
$\sigma\sim \dot{\gamma}^{0.4}$. The excitement in the field is due
to the fact that this is one of the first examples of how a simple
microscopic model leads to non-trivial rheological behavior at
mesoscopic or macroscopic scales, and the fact that this is
intimately related to the jamming scenario.

The more customary way to plot rheological data is to plot the stress $\sigma $ as a function of the imposed shear rate $\dot{\gamma}$; this is how data like that of Fig.~\ref{jamfig:10} are 
usually presented. At the time of writing this review, there is a
lot of debate about the precise exponents
\shortcite{olsson,hatano08a,hatano08b,hatano08c,olsson09,remmers,haxton09},
the existence of a quasistatic limit, the influence of the
mean-field approximation employed in the simulations of
\shortcite{olsson}, the question of whether there is a finite
velocity correlation length which diverges at the jamming point, or
one that is always of order the system size, etc. From our point of
view, the essential question however is whether, just like in the
static case, the exponents should be rational values which can be
understood in terms of simple arguments based on our insight in the
response of a marginal solid. To date, the answer to this question
is unknown.

 \section{Real physical systems} \label{jamsection:3.3}

 As stated in the introduction, space limitations do not allow us pay tribute appropriately to the recent experimental developments that provide new impetus to the field. Some of these are reviewed in the chapters by Cipelletti and Weeks on colloids and by Dauchot, Durian and van Hecke on grains and foams. We wish to draw attention to a few additional points here.

\subsection{Granular materials}

Excited granular matter turns out to be an excellent model system to probe dynamical heterogeneities and the pair-correlation function $g(r)$ --- think of the experiments with discs shaken on a table \shortcite{lechenault08a,lechenault08b} or the ball bearing experiments \shortcite{abatePRE06,keysNP07} discussed in section \ref{sectionstructural}. Unfortunately, it is extremely difficult to probe the vibrational density of states with grains. The scaling of static properties like the contact number is also difficult to study experimentally: in experiments one will always have friction, whereas most of the clean theoretical scaling predictions are for frictionless spheres --- as we have seen in section \ref{sectionfriction}, with friction the behavior depends stongly on the way a packing is made, so it is not even clear whether clear scaling behavior should be expected in the experiments on photoelastic discs \shortcite{behringer07} aimed at measuring $Z$ as a function of packing density (However, experiments using photoelastic discs have been very powerful in studying force-chain networks and correlations \shortcite{trushnature}.)  A powerful way to probe elasticity is to use ultrasound measurements, which, in agreement with what we have argued above, support a small $G/B$ ratio in sand \shortcite{bruno,jacob} and a sharp transition toward strong scattering \shortcite{jia} as the frequency increases. Another possibility which is being explored to probe the contact statistics and elastic moduli by using particles which swell when the humidity of the environment is changed \shortcite{Cheng09}.  Finally, experiments on avalanches are convenient for probing the length scales involved in failure and flows \shortcite{pouliquen}. In particular the angle at which an avalanche starts depends strongly on the thickness of the granular layer. A model based on the stabilization of anomalous modes by the rough fixed boundary at the bottom of the granular layer 
is consistent with the observed dependence
\shortcite{matthieuavalanche}.

\subsection{Emulsions and foams}

It is important to stress here that some of the observations
associated with the jamming approach --- in particular, the scaling
$\Delta Z\sim \sqrt{\Delta \phi}$ --- already appeared in some form
in the early work on bubbles and the bubble model
\shortcite{durian95,durian97}. Moreover,  it has become increasingly
clear that wet foams are a very good experimental model system for
jamming studies, since the friction is dynamic, not static. As a
result, models of frictionless spheres or discs provide a reference
frame to study the effect of shear. Indeed, recent experiments on
sheared foams \shortcite{katgert07,katgert09} hold the promise for
testing some of the predictions of the sheared bubble model that
were discussed in section \ref{jamsection:3.2}.

Another promising route is to use emulsions or foams. Here too,
experiments from over a decade ago
\shortcite{mason95emulsions,saintjalmes} already gave indications
for deviations from the behavior expected for affine deformations.
New, more-refined experiments, are called for. Emulsions have
recently also been used to probe the structure of static packings
\shortcite{clusel09}.

\subsection{Colloids} \label{sec3.4}

Soft colloids, such as foams, emulsions and particles made from poly(N-isopropyl acrylamide) microgels (NIPA particles)~\shortcite{NIPArefs1,NIPArefs2}, can be used to study densities above the jamming transition.  In aqueous solution, NIPA colloids swell substantially as the temperature is reduced only slightly.  As a result, the packing fraction of the sample can be varied over a wide range with only a minimal change of temperature.  The systems are thermal in that the particles are small enough (micron-sized) to exhibit significant Brownian motion.  Thus, such systems are well-suited for studying 
jamming at a fixed non-zero temperature.

\section{Connection with Glasses}
    In this section, we will describe some of the potential connections between the physics that appears at the jamming transition at point J and what happens in real molecular glass formers.  There are several phenomena that appear ubiquitously in amorphous solids and glasses yet are quite different from what is found in ordinary crystals.

\subsection{Properties of glasses to be considered}

    {\em Density of low-temperature excitations and the Boson peak:} There are many more low-frequency excitations in a glass than in a crystal.  One of the hallmarks of glasses are the so-called anomalous low-temperature excitations.  Since 1971 \shortcite{zeller71}, it has been known that the specific heat, $c_V$ of glasses varies approximately linearly with temperature rather than as $T^3$ as would be predicted by the Debye calculation where the low-frequency modes are long-wavelength planewaves\shortcite{AshcroftMermin}.  At a somewhat higher frequency, above a frequency $\omega_{BP}$, there is a dramatic excess of vibrational modes in glasses, known as the Boson peak\shortcite{sokolov}.  That the Boson peak should be a ubiquitous property of amorphous solids is surprising since the underlying structure, which would normally determine the phonon spectrum in a crystal, is so different in different glasses. The linear specific heat is surprising since even disordered systems should have well-defined elastic moduli when averaged over large enough wavelengths.  Therefore, at low enough frequencies, the modes should remain planewaves with a Debye spectrum.  This linear specific heat has been ascribed to the quantum tunneling of groups of atoms between roughly equivalent structural configurations\shortcite{AHV,Phillips72} although these configurations have not been unambiguously identified.

    {\em Thermal conductivity:}  In a related vein, the heat transport in glasses is decidedly different from what it is in crystals\shortcite{phillips_book}.   At very low temperatures, the thermal conductivity, $\kappa \sim T^2$ rather than $\kappa \sim T ^3$ as in crystals.  At higher temperatures there is then a plateau region. Above the plateau there is a third regime where $\kappa$ increases gently, $\kappa \sim T$.  This behavior is surprisingly universal between different glasses yet there is no generally-accepted explanation of either of the higher-temperature regimes.  The lowest-temperature region, where $\kappa \sim T^2$, has been modeled as the scattering of plane waves off the localized tunneling systems used to explain the low-temperature linear term in the specific heat.

    {\em Failure under applied stress:}  When an amorphous solid is compressed or sheared sufficiently, it will start to fail.  However, the failure often occurs in the form of rather localized rearrangement events rather than via a long-range catastrophic collapse\shortcite{langer98,langer08,harrowell}.   Since the system goes unstable when the frequency of a mode reaches zero, it is possible to identify a rearrangement with the lowest-frequency mode just before the rearrangement~\shortcite{maloney}.  However, ideally one would like to predict well in advance where failure will occur, even in large systems with many plane-wave-like modes at low frequencies.

    {\em Structure: } In metallic and colloidal glasses, the structure has often been described in terms of sphere packings\shortcite{Finney,Cahn,Zallen}.  While this appears to be quite natural, there are certain features that have been related to the glassy structure that are not well understood.  For example, the first peak in the pair distribution function, $g(r)$ is tall and sharp and there is a split second peak.  Moreover, as the temperature is lowered in a supercooled liquid, there is no sign of the onset of rigidity in the static structure factor, $S(q)$;  the first peak in $S(q)$ simply grows smoothly in height and decreases in width as the temperature is lowered through the glass transition temperature, $T_g$\shortcite{Busse}.

\subsection{Relating the Jamming Paradigm to Glasses}

{\em Low-temperature excitations and the Boson peak in covalent
glasses}

\begin{figure}[t]
\centering \vspace{-0.0 cm}
\includegraphics*[angle=0,width=.97\textwidth]{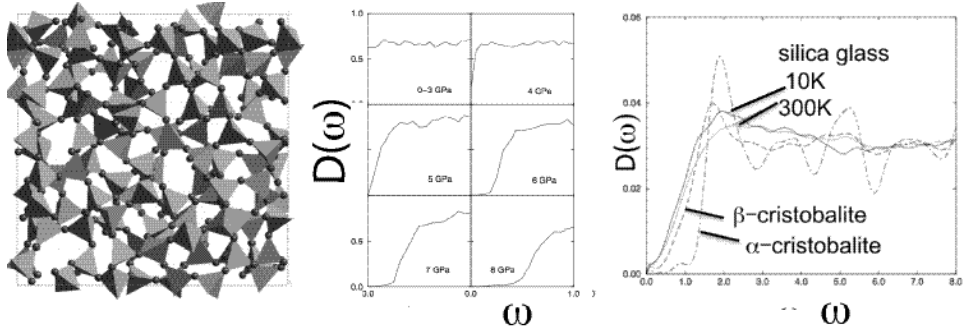}
\vspace{-0 cm} \caption[]{Left:  Configuration of amorphous silica
as modeled  by RUM:  SiO$_4$ tetrahedra are connected by springs of
finite stiffness at the joints, which can freely rotate
\protect\shortcite{silica1}.  Center: Density of states $D(\omega)$
for amorphous silica configurations obtained at various pressure
within the RUM approximation. At low-pressure, there are only
tetrahedra, and $D(\omega)$ is essentially flat. As the pressure
increases 5-fold coordinated Silicon atoms become more and more
numerous, and $D(\omega)$ erodes at low-frequency due to increased
coordination, very much like in sphere packings
\protect\shortcite{silica2}. Right: Computations of $D(\omega)$ for
atomic models of silica. Silica glass, $\beta$-cristobalite and
$\alpha$-cristobalite all present a plateau above 1 Thz, the Boson
peak frequency \protect\shortcite{silica1}. } \label{figsilica}
\end{figure}

Various  theoretical approaches  can reproduce a Boson peak and 
some of the features of the corresponding normal modes. They are
based on disorder and generally connect the peak to an elastic
instability, as in effective medium models
\shortcite{schirmacher,schirmacher2}, euclidian random matrix theory
\shortcite{grigera3} or mode coupling \shortcite{mayr}.  However,
the structural parameters controlling the peak are imposed by hand
in these theories (amount of disorder in \shortcite{schirmacher2},
density in \shortcite{grigera3} or structure factor in
\shortcite{mayr}). What aspects of the structure are the most
relevant for  the peak is unclear.
Moreover, disorder cannot be a generic explanation for the Boson peak: silica 
has one of the largest peaks, but so do the corresponding crystals,
the cristobalites, as shown in Fig.~\ref{figsilica}. It is sometimes
argued that silica is a special case. The variational argument of
section \ref{section1.2.3} shows that  is not so: generically
amorphous and crystalline elastic networks must have qualitatively
similar low-frequency spectra if their coordination is identical.
For example, a cubic lattice has a flat spectrum as does a random
close packing of elastic particles.
 Thus coordination rather than positional disorder matters (but disorder in the spring stiffnesses obviously matters and can be incorporated in an improved variational argument \shortcite{wyartthesis}).

The variational argument  of \shortcite{wyartEPL} has broad
applications, as illustrated for silica and other tetrahedral
networks following the analysis of \shortcite{wyartthesis,matsi}.  A
particularity of silica or germanium oxide is the weak force
constant associated with the rotation of two adjacent tetrahedra, in
comparison with deforming the tetrahedra themselves. This suggests
modeling silica as an assembly of stiff tetrahedra connected by
flexible joints, the RUM model \shortcite{silica1}, as sketched in
Fig.~\ref{figsilica}.  When this approximation is used on
configurations obtained at various pressures to compute the
vibrational spectrum, one finds a remarkable similarity between
silica and particles near jamming as illustrated in
Fig.~\ref{figsilica}. The cause is identical: at reasonable
pressures, silica is made of tetrahedra, and 5-fold defects are
extremely rare. If joints are flexible, a tetrahedral network is
isostatic (there are 6 degrees of freedom per tetrahedron, and 3
constraints per joint, and twice as many joints than tetrahedra) and
must have a flat density of states, as observed. For real silica,
the rotation of the joints has a finite stiffness, which must shift
the anomalous modes forming the plateau by some frequency scale
$\Delta$. Using {\em ab initio} computation of the joint-bending
force constant, and the molecular weight of silica one estimates
$\Delta=1.4 Thz$. Empirically the density of states of silica indeed
has a plateau above approximatively 1 Thz, as expected.   This is
shown in Fig.~\ref{figsilica} for numerical simulations. This
argument also applies to germanium oxide. For amorphous silicon, the
bending force constant of the joints is comparable to the other
interactions, since they correspond to the same covalent bond, and
one therefore does not expect a Boson peak at low frequency; indeed
in this material the Boson peak is nearly non-existent. Ref.
\shortcite{wyart08} extends these arguments to other covalent
networks.

{\em Low-temperature excitations and the Boson peak in molecular
glasses}

We can make progress in understanding the excess number of
low-frequency excitations in molecular glasses by recalling some of
the properties of solids near the jamming transition.  At point J,
the jamming transition for frictionless repulsive spheres, the
density of states has a plateau all the way down to zero frequency.
Upon compression, the plateau persists only down to a lower cutoff
frequency, $\omega^*$\shortcite{silbertPRL05,epitome}
[Figs.~\ref{jamfig:2} and \ref{jamfig:3}(b)] as explained in
\shortcite{wyartEPL}.  Below that frequency, the normal modes are
weakly scattered plane waves while above it they are strongly
scattered\shortcite{Silbert09,Xu09,Vitelli09,matthieucpap} .  This
jump in $D(\omega)$ appears to be the counterpart to the dramatic
increase in the number of vibrational modes that appear at the Boson
peak in glasses\shortcite{silbertPRL05,Xu07,wyartthesis,wyartEPL} .

In order to assert this correspondence more generally, we have
ascertained that the physics dominating point J for repulsive
spheres is also operative for other systems such as those with
particles of non-spherical shape or with particles mediated by
long-range and/or attractive interactions.  To include such effects,
we have studied several different models of glasses that can be
analyzed in terms of their proximity to the jamming transition. We
refer to section \ref{jamsection:3.1} on ellipsoids for a brief
discussion of what happens when the particles are non-spherical  ---
in essence, the results are consistent with the  jamming scenario
based on the analysis of frictionless spheres, though in a
surprising and nontrivial way.

We have also studied models that were based on variants of the
Lennard-Jones interactions between spheres.  Such potentials allow
attractive as well as long-range interactions so that the average
coordination number can be arbitrarily large.  Lennard-Jones
potentials are more realistic models of molecular systems than the
simple harmonic force law described above because the potential
decays rapidly with inter-particle separation.  Particles separated
by even a slightly larger distance, interact much more weakly than
those closer together.

We have shown  \shortcite{wyartthesis,Xu07} that the onset of excess
modes in these systems derives from the same variational-argument
considerations that arise at point J for systems with finite-ranged
repulsions.  It can again be estimated by analyzing the vibrational
energy originating from the excess contacts per particle over the
minimum number needed for mechanical stability: $Z - Z_c$.  The
extra contacts can be divided into $Z_1$ ÒstrongÓ contacts and the
remaining $Z-Z_1$ ÒweakÓ ones.  Strong contacts shift the energy
cost of a mode that is initially zero in the isostatic limit to a
nonzero value according to the variational argument outlined above,
while weak contacts shift the energy simply by increasing the
restoring force for displacements.  The energy is then minimized to
obtain the fraction of strong contacts.  Even though these glasses
have a high coordination number, most of the additional contacts can
be considered to be weak.

\begin{figure}[t]
\centering
\includegraphics*[width=.98\textwidth]{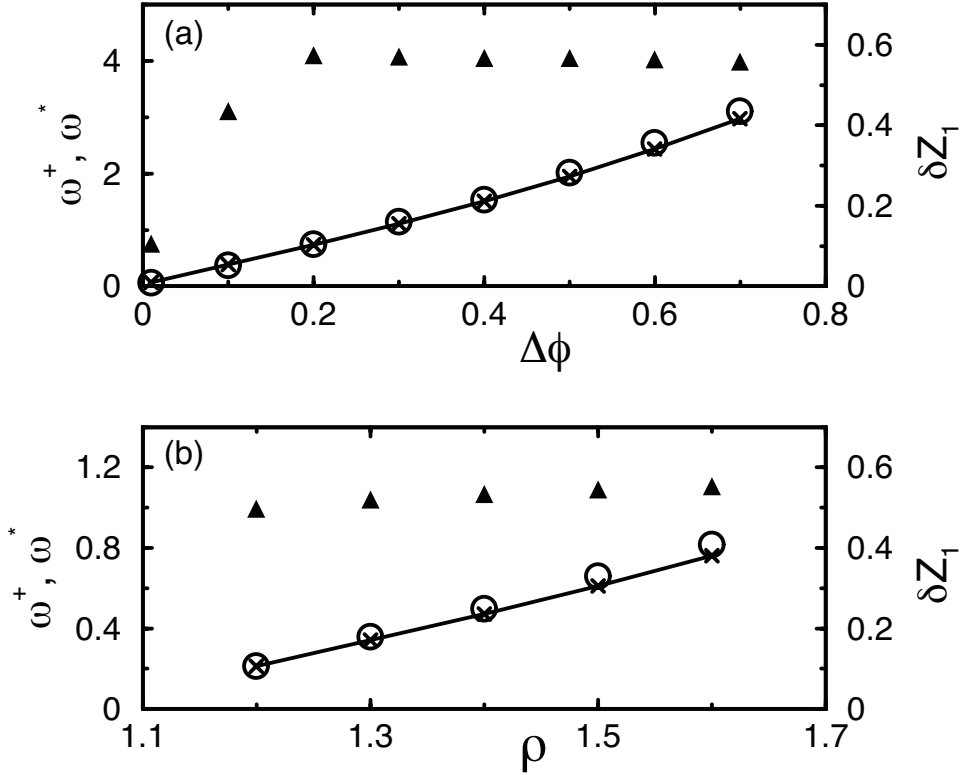}
\caption[]{Characteristic frequencies and fractional deviation from
isostaticity for two three-dimensional systems based on
Lennard-Jones potentials.  Left panel shows data for repulsive
Lennard-Jones bidisperse mixtures where only the repulsive part of
the potential is included versus $\Delta \phi$, distance from the
unjamming transition.  Right panel shows data versus density,
$\rho$, for a full Lennard-Jones potential with cutoff at
$r=2.5\sigma_{i,j}$ where $\sigma_{i,j}$ is the average diameter of
particles $i$ and $j$.  On the left axis, $\omega^{\dagger}$ (open
circles) is the onset frequency of the anomalous modes calculated
from numerical simulations and the corresponding predictions for
$\omega^*$ (crosses) based on a variational calculation.  The
agreement is very good.  Lines are to guide the eye.  On the right
axis is the fractional deviation from isostaticity: $\delta Z \equiv
(Z_1 - Z_c)/ Z_c$ (solid triangles). Figure taken from
\protect\shortcite{Xu07}} \label{jamfig:15}
\end{figure}

These results are shown in Fig.~\ref{jamfig:15} for two models.
$\omega^*$ is the theoretical prediction for the onset frequency of
excess modes based on the variational argument.  The onset frequency
determined from simulation is given by $\omega^{\dagger}$.  The
agreement between $\omega^*$ and $\omega^{\dagger}$ is very good.
On the right-hand axis $\delta Z \equiv (Z_1 - Z_c)/ Z_c$ is shown.
Note that even though $Z$ itself can be arbitrarily large, $\delta
Z$ remains small and does not get larger that $0.6$.  This is why
these systems with high coordination can still be understood in
terms of the physics of point J and isostaticity.  Even though it
may be impossible to reach the jamming transition itself Ñ for
example because there are long range interactions or attraction (as
in the Lennard-Jones system) Ñ the effects of the long-ranged part
of the potential can be treated as a correction to the predominant
effects of jamming.

{\em Thermal conductivity:} The thermal conductivity, $\kappa$, of a
system near the jamming threshold has been calculated
and captures well the thermal conductivity of silica glass above its
plateau\shortcite{Xu09,Vitelli09}. Because the phonons are scattered
so strongly by the disorder, phonon-phonon scattering is relatively
unimportant so the thermal transport can be calculated within the
harmonic approximation without recourse to the anharmonic properties
of the modes.  This is in strong contrast to the situation for
crystals, where anharmonicity must be included to obtain a thermal
conductivity that is not infinite.  At the jamming threshold, the
long-wavelength phonons that give rise to a diverging thermal
conductivity in the harmonic limit are completely suppressed by the
excess modes, since the density of vibrational states remains flat
down to zero frequency.  As a result, the thermal conductivity is
finite even within the harmonic approximation at this special point.
The thermal conductivity can be expressed in terms of the
diffusivity, $d(\omega)$ (defined in section
\ref{sectiondiffusion}), and heat capacity, $C(\omega)$, of each
mode:
\begin{eqnarray}
\kappa(T)&=&\int_{0}^{\infty} {D}(\omega)  \!\!\!\! \quad C(\omega)
{d}(\omega) \!\!\!\! \quad d\omega \!\!\! \quad ,
\label{eq:conductivity}
\end{eqnarray}

In a weakly scattering system such as a crystal, $d(\omega) =
c\ell(\omega)/3$ where $c$ is the sound speed and $\ell(\omega)$ is
the phonon mean-free path.  As discussed in section
\ref{sectiondiffusion}, Fig.~\ref{jamfig:7} shows that at the
jamming transition $d(\omega)$ has a very simple shape: it is a
plateau that extends from our lowest frequency up to the onset of
the high-frequency localized states \shortcite{Xu09,Vitelli09},
where $d(\omega)$ falls rapidly to zero.  The magnitude of the
diffusivity in the plateau region, $d_0$, is small and scales as:
$d_0 \sim \sigma^2 \sqrt{\frac{k_{eff}}{M}}$ where $k_{eff} =
\frac{\partial^2 V(r_{ij})}{\partial r_{ij}^2}$ is the effective
spring constant of the system that scales as the bulk modulus, $B$.
Effective medium theory suggests that these results hold more
generally for weakly-coordinated disordered elastic networks
\shortcite{matthieucpap}.

Upon compression, the plateau region in $d(\omega)$ no longer
extends down to zero frequency but only to a frequency that is
proportional to $\omega^*$.  Thus all the modes above $\omega^*$
(excluding the truly localized modes at high frequency) have a
nearly constant diffusivity.  As we emphasized in the section on
low-temperature excitations, $\omega^*$ obtained from the density of
states corresponds to the frequency of the Boson peak.  Here we find
that this same frequency also is proportional to the onset of a
region of flat diffusivity for thermal
transport\shortcite{Xu09,Vitelli09}.  It is just such a region of
constant diffusivity that was postulated by Kittel\shortcite{kittel}
to explain why the thermal conductivity of glasses has a weak,
nearly linear, temperature dependence above the thermal-conductivity
plateau.  This is consistent with other evidence in glasses that the
end of the plateau in $\kappa (T)$ corresponds to the Boson
peak\shortcite{zeller71}.

{\em Failure under applied stress:}  As summarized in section
\ref{sectiondiffusion}, the vibrations of soft-sphere packings are
quasi-localized near the onset of anomalous modes, $\omega^*$: they
have large displacements of relatively few particles while the rest
of the mode has a very small amplitude.  As the sample is compressed
or sheared, it will eventually go unstable and rearrange into a new
configuration.  At zero temperature, this instability is governed by
some quasi-localized mode in the system that goes ``softÓ so that
its frequency is pushed down until it reaches
zero~\shortcite{maloney}. At this point, the sample moves into a
new configuration.    One cannot predict too much from the purely
harmonic properties of a mode about what it will do when it is
pushed to such a large amplitude that is goes unstable.   However,
at the very least one can say that at the point of instigation, the
instability will initially have a very low participation ratio. This
is consistent with the observation that when a glass fails due to
pressure or shear stress, the failure tends to be highly localized
in shear-transformation zones~\shortcite{langer98,langer08}. So far,
however, it has not been possible to identify shear-transformation
zones unless they flip (unless a rearrangement occurs).  The
existence of many low-frequency quasilocalized modes with low energy
barriers suggests that it may be possible to identify multiple
shear-transformation zones from structural properties of the
quasilocalized modes, and to predict which ones will flip {\it a
priori}, even in large systems where there are many plane-wave-like
modes at low frequencies.


    {\em Structure:}  At point J, the pair distribution function has many singular properties.  In particular, we have shown that the first peak in $g(r)$ is a $\delta$-function at the nearest neighbor distance, $\sigma$.  The area under this $\delta$-function is just the isostatic coordination number: $Z = 6$ in three dimensions.  On the high side of this $\delta$-function, $g(r)$ has a power-law decay: $g(r) \propto (r-\sigma)^{-0.5}$.  Also, at the jamming transition, the second peak of $g(r)$ is split into what appear to be two singular sub-peaks Ð one at $r= \sqrt {3}\sigma$ and the other at $r=2\sigma$.  There is a divergent slope just below each subpeak and a step-function cutoff on its high side.  The existence of a split second peak in $g(r)$ is a feature that is seen in many glasses, such as metallic and colloidal glasses.   One can thus trace the origin of these subpeaks to the geometry at the jamming threshold.  One can also ask why, as a supercooled liquid is cooled into the glassy state, there is no signature in the static structure factor, $S(q)$, which is how the structure of molecular glasses is determined from scattering experiments.  At the jamming threshold, there are indeed clear signatures in the structure of a sample: these are the divergences just mentioned in $g(r)$.  However, because $S(q)$ is the Fourier transform of $g(r)$, the divergences simply appear as slightly more pronounced oscillations in $S(q)$ out to very large wavevector.


\section{Connection with Super-cooled liquids}


The understanding of the glass transition remains one of the
enduring mysteries of condensed matter physics.  There is no doubt
that the time scale for relaxation increases dramatically (faster
than an Arrhenius law for what are called ``fragile" glasses) as the
temperature is lowered from the melt towards the glass transition
temperature, $T_g$, but it is much less clear whether there are any
static growing length scales that can be identified that accompany
the slowing down\shortcite{Ediger,Debenedetti}.   However, as is
well-documented in this book, there is a growing length scale
associated with heterogeneities in the dynamics, which become
increasingly collective as the viscosity increases.  The origin of
the super-Arrhenius increase of relaxation time with decreasing
temperature ({\it i.e.} fragility) is controversial.  Is it due to
free-volume effects, thermal activation over energy barriers that
grow with decreasing temperature, or some other type of cooperative
motion?

\subsection{Glass transition and soft modes in hard sphere liquids}\label{HS}



Goldstein \shortcite{goldstein} has proposed that the slowing down
of the dynamics is associated with the emergence of meta-stable
states as a liquid is cooled.  Such a roughening of the energy
landscape is predicted by mean-field models of liquids and spins
\shortcite{mezard,kurchan,castellani,bb}  at the so-called
mode-coupling temperature $T_{MCT}$.  However these theoretical
approaches lead to a (non-observed) power-law divergence of the
viscosity at $T_{MCT}$, and their interpretation in real space is
unclear and actively studied \shortcite{franz,biroli}. Empirically,
a direct validation of Goldstein scenario in liquids is difficult.
Several numerical analysis\shortcite{angelani,giardina,grigera} have
claimed to observe a transition in the energy landscape at
$T_{MCT}$, but in our opinion the interpretation of these results is
disputable \footnote[1]{ In these numerics,  equilibrium
configurations close to the configurations visited by the dynamics
are analyzed. Their number of saddles (or unstable modes) is
computed in function of the temperature, and it is argued that this
quantity can be fitted by a curve that vanishes at a temperature
$T_{MCT}$, defined here by fitting the dynamics as $\tau\sim
(T-T_{MCT})^{-a}$, where $\tau$ is the $\alpha$-relaxation time
scale and $a$ some exponent. From this observation it is stated that
meta-stable states appear at $T_{MCT}$. However, this observation is
precisely the behavior expected if meta-stable states had already
appeared at larger temperature, implying that the dynamics is
activated in the neighborhood of $T_{MCT}$. Indeed assuming
activated dynamics,  the number of saddles $n_s$  reflects the
probability that a region is in the process of rearranging (crossing
a barrier), which occurs with a probability of order $1/\tau$. Thus
one expects $n_s\sim 1/\tau$, which can be fitted with a vanishing
quantity at  $T_{MCT}$ simply due to the definition of that
temperature. }. Another intriguing aspect of Goldstein proposal is
its application to hard spheres, where free volume rather than
energy matters.

\begin{figure}[t]
\centering \vspace{-2 cm}
\includegraphics*[angle=0,width=1\textwidth]{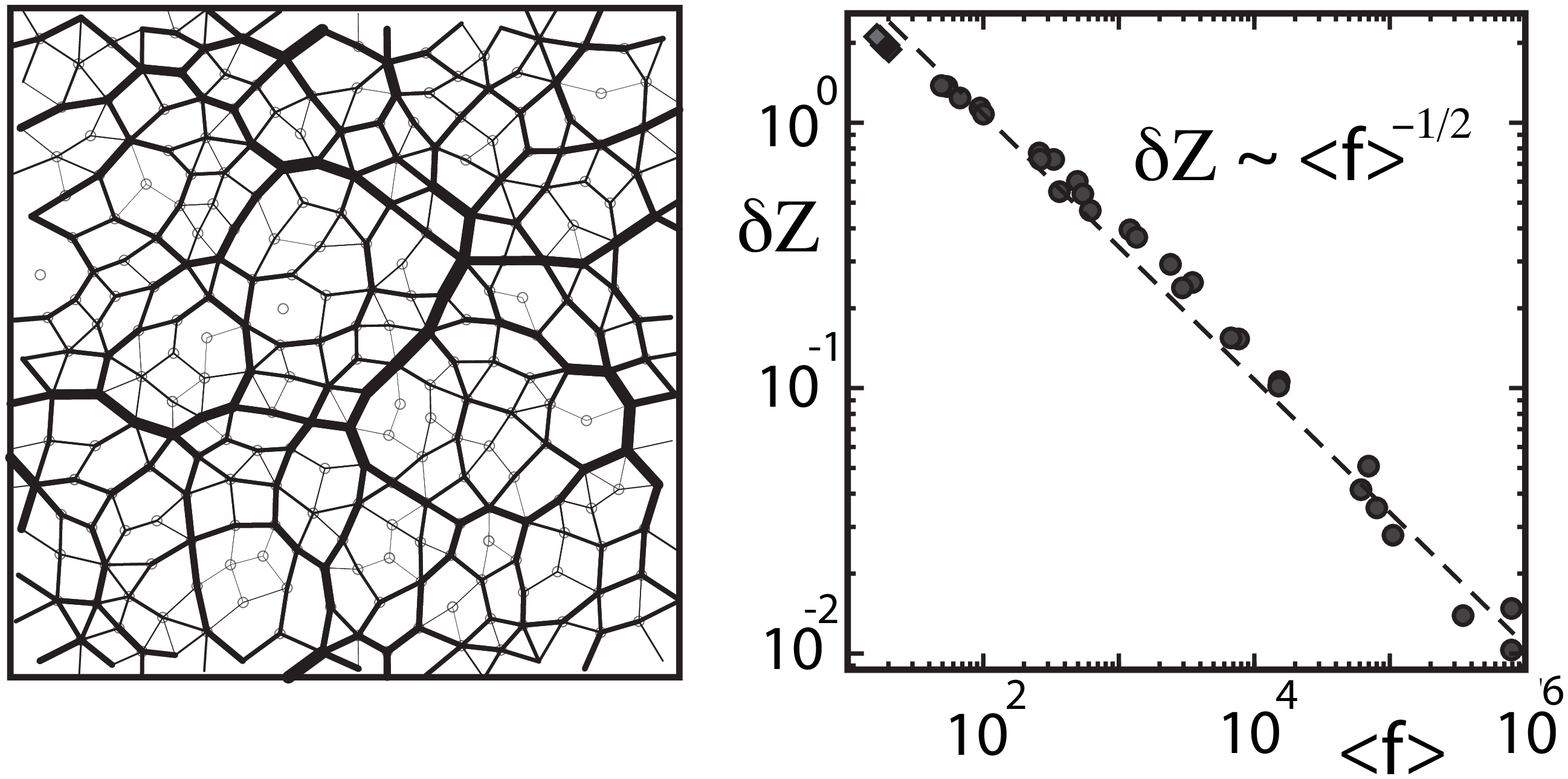}
\vspace{-1.5cm} \caption[]{Left: Illustration of the fact that the
problem of hard elastic discs at finite temperature and density in
the supercooled liquid or glass phase (in this case $\phi_c-\phi=
2\cdot 10^{-4} \phi_c$), can, in between rearrangement events, be
mapped onto the problem of a static packing with a logarithmic
interaction  potential  \protect\shortcite{brito07a,brito07b}.
Essentially, particles experience a reduced free volume due to
collisions with close neighbors, leading  to an effective repulsive
interaction. Thus particles exert a force on each other. The plot
shows the average contact forces for 256 particles averaged over
$10^5$ collisions. The points represent the centers of the
particles, and the thickness of the line is proportional to the
force strength. Since in between rearrangement events the structure
is stable, forces are balanced on each particle
\protect\shortcite{brito07a,brito07b}. Right: $\Delta Z$ inferred
from the coordination of the force network {\it vs}. average contact
force $\langle f \rangle$, a measure of pressure. Measurements are
made in the supercooled liquid (diamonds) and in the glass
(circles). The observed coordination is well-captured by its minimal
value $\Delta Z\sim \langle f \rangle^{-1/2}$ allowing mechanical
stability, supporting that the glass is marginally stable
\protect\shortcite{brito3}.} \label{jamfig:13}
\end{figure}

The connection between microscopic structure and vibrational
spectrum established above directly supports that the emergence of
meta-stable states slows down the dynamics, and yields a geometrical
interpretation of this phenomenon \shortcite{brito3}.  In order to
see that, the theoretical description of anomalous modes and their
associated length scales must be extended to finite temperature. An
analogy between hard-sphere liquids and athermal elastic systems can
be built \shortcite{brito07a,brito3} in two steps:  {\em (i)} on
short time scales particles rattle rapidly but no large
rearrangements occur. As illustrated in Fig.~\ref{jamfig:13}, a
contact force network can be defined, following earlier work on
granular matter \shortcite{bubble66}, which represents the average
force exchanged between particles.  From such networks a
coordination number is computed by counting the pairs of particles
with a non-zero contact force. {\em (ii)} All the possible
hard-sphere configurations associated with a given network can be
summed up, leading to a computation of the free energy that can be
expressed in terms of the mean positions of the particles. The
effective interaction is found to be logarithmic. Expanding the free
energy defines a dynamical matrix and vibrational modes that
characterize the linear response of the mean particles positions to
any imposed external forces within a meta-stable state.

The results on the vibrational spectra of athermal elastic networks
apply to this case as well, and the stability criterion of
Eq.~(\ref{maxext}) must be satisfied in any metastable
configuration. For hard spheres, it can be written $\Delta Z >
p^{-1/2}\sim h^{1/2}$ as the pressure $p$ satisfies $p\sim
1/h\sim1/\Delta \phi$, where $h$ is the typical average gap among
particles in contact. As illustrated in Fig.~\ref{jamfig:13}, the
bound is approximatively saturated and the glass appears to have
just enough contacts to maintain its stability.  Marginal stability
is a fundamental feature of the glass, generically foreign to
crystals. It also occurs in some mean field spin models
\shortcite{kurchan}. In hard spheres it implies the presence of
anomalous modes near zero frequency and a curious behavior of the
short-time dynamics, in particular a dependence on the mean-square
particle displacements scaling as  $\langle \delta R^2\rangle \sim
\Delta \phi^{3/2}$ rather than the $\Delta \phi^2$ dependence
expected in a crystal \shortcite{brito3}. Those predictions have
been confirmed numerically \shortcite{brito07b,brito3,Kurchan2}.

\begin{figure}[t]
\centering
\includegraphics*[width=.6\textwidth]{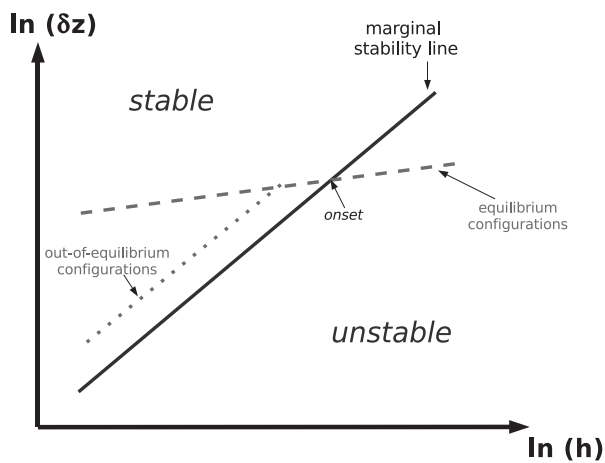}
\caption[]{Phase diagram for the stability of hard-sphere
configurations, in the coordination $\delta z$ {\it vs} average gap
$h$ plane. The marginal stability line  delimits stable and unstable
configurations. The dashed  line correspond to equilibrium
configurations for different $\phi$. As $\phi$ increases, $h$
decreases and the two lines eventually meet. This occurs at the
onset packing fraction $\phi_{onset}$, where the dynamics become
activated and thus intermittent. At larger $\phi$, viscosity
increases sharply as configurations visited become more stable. For
a finite quench rate the system eventually falls out of equilibrium
at the glass transition packing fraction $\phi_0$. More stable and
more coordinated regions cannot be reached dynamically, and as
$\phi$ is increased further, the system lives close to the marginal
stability region, as indicated in the dotted line. The location of
the out-of-equilibrium trajectory depends on the quench rate. In the
limit of very rapid quench, the out-of-equilibrium line approaches
the marginal stability line \protect\shortcite{brito3}.}
\label{explanation}
\end{figure}

Why is a hard sphere glass marginal?  It has been argued
\shortcite{brito3} that this must be so if the viscosity increases
very rapidly when meta-stable states appear in the free-energy
landscape. In the  $(\Delta Z, h)$ plane sketched in
Fig.~\ref{explanation}, Eq.~(\ref{maxext}) draws a line separating
stable and unstable configurations.  At equilibrium the mean value
of  $\Delta Z$ and $ h$ are well-defined functions of $\phi$, and
the corresponding curve is represented by the dashed line in
Fig.~\ref{explanation}.  At low $\phi$, gaps between particles are
large and the configurations visited are unstable. As $\phi$
increases, the gaps decrease and configurations eventually become
stable. This occurs at some $\phi_{onset}$ when the equilibrium line
crosses the marginal-stability line.  At larger $\phi$, the dynamics
becomes activated and intermittent, and the viscosity increases
rapidly according to our hypothesis, so that on accessible time
scales equilibrium cannot be reached deep in the regions where
metastable states are present.  As a consequence, the system falls
out of equilibrium at some $\phi_0$ larger than  $\phi_{onset}$.
Configurations visited must therefore lie close to the marginal
stability line, as represented by the dotted line in
Fig.~\ref{explanation}, since more stable, better-coordinated
configurations cannot be reached dynamically.

\begin{figure}[t]
\centering \vspace{-1.5cm}
\includegraphics*[width=0.7\textwidth]{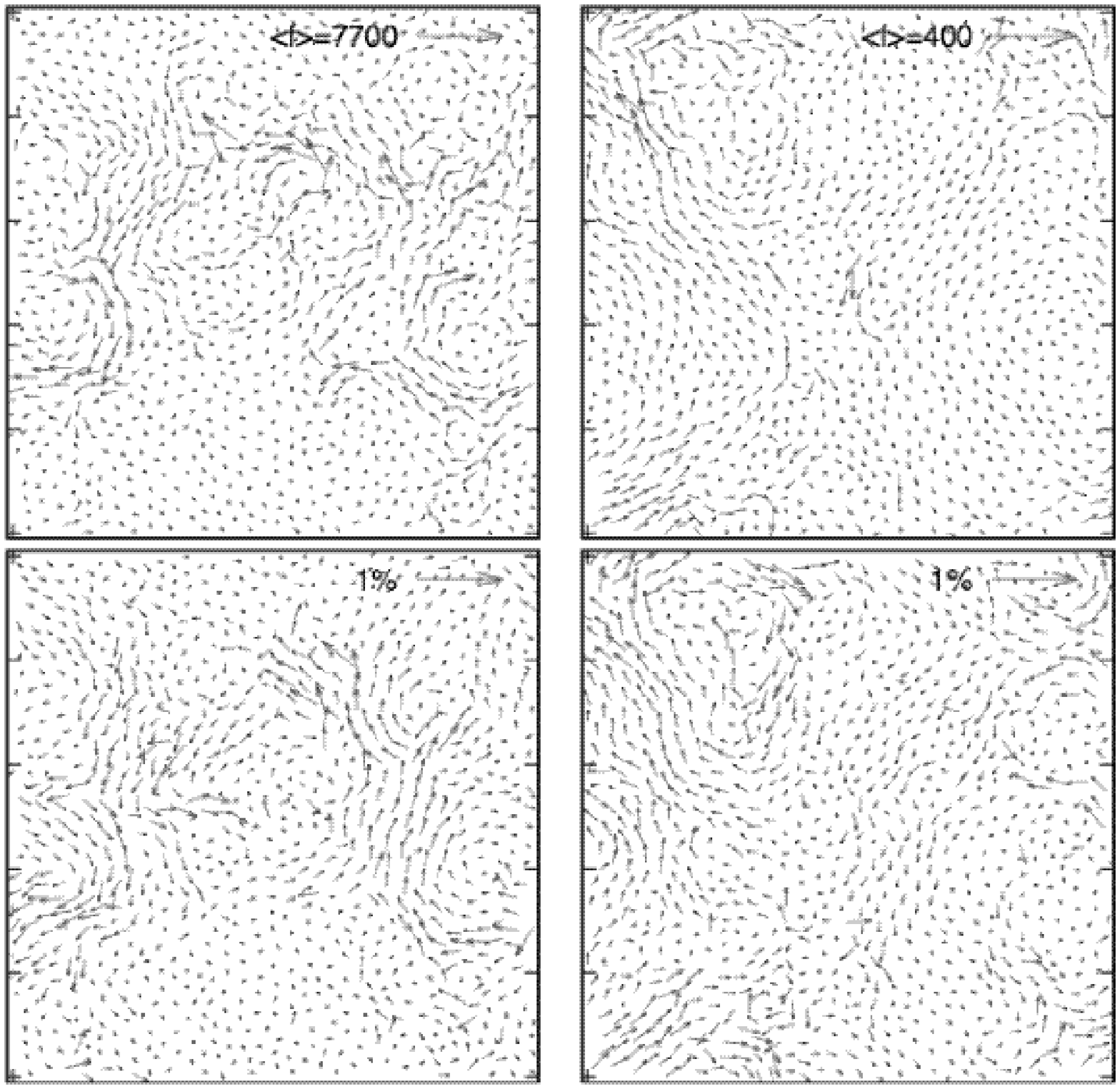}
\vspace{0 cm} \caption[]{Above: two examples of sudden
rearrangements in the glass phase for different average contact
force $\langle f \rangle$ for $N=1024$ particles.  Below: projection
of these rearrangements on the  $1\%$ of the normal modes that
contribute most to them, which systematically lie at the lowest
frequencies \protect\shortcite{brito3}.  This projection captures
most of the displacement, indicating that a limited number of
(collective) degrees of freedom are active during rearrangements.}
\label{relaxation}
\end{figure}

Further support for the view that the viscosity is governed by the
presence of an  elastic instability near $\phi_{onset}$ comes from
the observation that the sudden rearrangements that relax the
structure occur along its softest modes (both in the liquid and in
the glass phase) \shortcite{brito07b}, indicating that the lowest
activation barriers lie in these directions, as illustrated in
Fig.~\ref{relaxation}.  Experiments in granular materials
\shortcite{brito4} and numerical observations of normal modes of the
energy of inherent structures
\shortcite{Schober2,harrowell,harrowell2} reach similar conclusions.
The normal modes of the free energy considered here are expected to
correlate with the dynamics much better than those of the energy,
and  must be considered  in particular when the interaction
potential is strongly non-linear. The collective rearrangements
involving a few tens of particles commonly seen in super-cooled
liquids correspond mostly to a few modes \shortcite{brito07b}, an
effect even stronger in the glass phase. This analysis supports that
the anomalous modes are the elementary objects to consider to
describe activation, and that the collective aspect of rapid
rearrangements is already present at the linear level in the
structure of the modes.

This approach permits  a quantitative study of activation, in particular to compute  the fraction of the spectrum contributing to  rearrangements. This  quantity decreases rapidly with compression near the glass transition \shortcite{brito07b}, implying that fewer and fewer degrees of freedom are relevant for the dynamics.  This observed reduction of active degrees of freedom presumably mirrors the slowing down and increasingly collective aspect of the rapid rearrangements that relax the structure \shortcite{candelier,candelier2}, but this has yet to be clarified. 

Overall,  the present approach furnishes a simple spatial picture
for the slowing-down of the dynamics at intermediate viscosities:
meta-stable states appear when  the contact force network become
sufficiently coordinated to resist the destabilizing effect of
pressure.  The localization for long times of an individual particle
is thus not due to some specific properties of the cage formed by
its neighbors, but has to do with the structure of the packing on a
length $l^*$ that can be large. Incidentally the concept of caging,
commonly used to picture dynamical arrest, is misleading.





 \subsection{Models of soft finite range repulsions at finite temperature}

By analyzing several models near the jamming threshold, some
universal aspects of the slow-relaxation dynamics that appear in one
asymptotic limit become apparent\shortcite{XuHaxton}.  In
particular, when the pressure, $p$, is small all the relaxation time
data can be suitably plotted on a single master curve as a function
of a single variable.

The models considered here all have finite-range repulsions.  For
this study, thermal energy was included in the simulations, and the
relaxation time $\tau$ of the system was measured by studying when
the intermediate scattering function, $f(q,t)$ drops to $e^{-1}$ of
its initial value.

\begin{figure}[t]
\centering
\includegraphics*[width=0.995\textwidth]{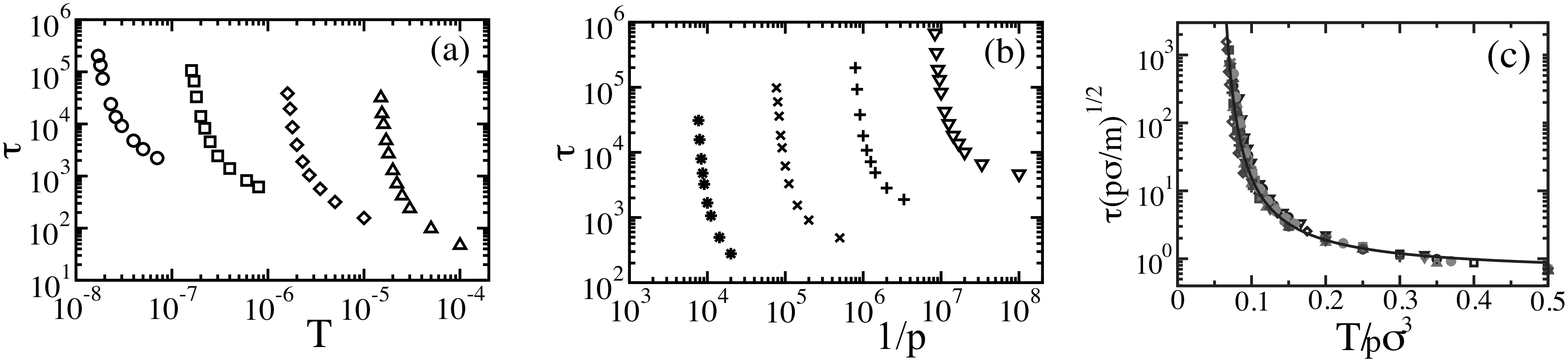}\hspace*{3mm}
\caption[]{Relaxation time, $\tau$, versus temperature, $T$, (a) and
versus inverse pressure, $1/p$, (b) for a system of spheres
interacting via repulsive harmonic potentials.  Panel (c) shows that
all the relaxation time data for different potentials (harmonic,
Hertzian and hard sphere) can be collapsed onto a single master
curve.  To get this data collapse, the relaxation time is
non-dimensionalized by pressure and plotted versus $T/p\sigma^{3}$.
Figure taken from \protect\shortcite{XuHaxton}. } \label{jamfig:14}
\end{figure}

Normally in simulations of this type, one measures time in units of
$\sigma \sqrt{m/\epsilon}$ where $\sigma$ is the particle diameter,
$m$ is the mass of the particle, and $\epsilon$ is the energy scale
of the potential.  However, in order to produce a collapse onto a
master curve, $\tau$ is non-dimensionalized in another way by using
$\tau \sqrt {p \sigma/m}$.  Plotting this quantity versus $T/(p
\sigma^3)$ produces an excellent collapse of the data for several
different potentials of interaction at all values of $T/(p
\sigma^3)$ as long as the non-dimensionalized pressure,
$p\sigma^3/\epsilon$ is sufficiently small\shortcite{XuHaxton}. This
is shown in Fig.~\ref{jamfig:14}.  This limit is satisfied at low
pressures in soft-sphere systems and at any pressure for a
hard-sphere system where $\epsilon \rightarrow \infty$.
Fig.~\ref{jamfig:14} shows that low-pressure soft-sphere data do
indeed collapse with the hard-sphere result.   In addition,
Ref.~\shortcite{XuHaxton} found that limiting hard-sphere behavior
remains a good approximation up to higher values of $p
\sigma^3/\epsilon$ as $T/(p \sigma^3)$ increases.

This data collapse has two important implications. First, it shows
that at sufficiently low pressures the relaxation at the glass
transition is a function of only a single variable.  The molecular
glass transition is, in this limit, equivalent to the hard-sphere
colloidal glass transition.  Second, Ref.~\shortcite{XuHaxton} shows
that there are systematic deviations from this universal behavior at
larger values of the second control parameter: $p\sigma^3/\epsilon$.
This demonstrates that there are at least two distinct physical
processes that enter to produce glassy dynamics. What is important
is that they can be cleanly separated, so that for sufficiently
small $p\sigma^3/\epsilon$, the rescaled relaxation time is only a
function of a single variable, $T/(p \sigma^3)$.   Note that the
existence of such a scaling function still does not tell us whether
the rescaled relaxation time diverges at $T/(p \sigma^3)>0$,
implying that a thermodynamic glass transition exists, or at $T/(p
\sigma^3)=0$, implying that the glass transition is an isostatic
jamming transition.

In order to understand the consequences of this result for molecular
liquids, we must first understand the corrections to the leading
hard-sphere behavior when we slowly increase $p\sigma^3/\epsilon$.
These studies, have been done for repulsive spheres near the jamming
threshold.  However, molecular liquids typically have attractions
and are at much higher densities than those considered here.  To
make the correspondence more general, we would have to push these
systems farther from the jamming threshold. Nevertheless, this
result shows how certain aspects of jamming, inherent in the
hard-sphere liquid, are important for glass-forming liquids
generally.  As we have shown, at least one other distinct
contribution to the relaxation must also be considered as the
pressure is increased.

As one increases $p\sigma^3/\epsilon$, there are corrections to the
leading hard-sphere behavior.  It has been argued that these
corrections can be collapsed onto a single curve by rescaling the
pressure by some factor that varies with packing
fraction~\shortcite{BerthierWitten}.

\section{Outlook to the future - A unifying concept}

The vibrational spectrum of crystals consists of plane waves and
leads to a description in terms of linear elasticity.  In those
systems, one can compute the non-linearities of the excitations and
how they scatter from defects.  Such a systematic description is
lacking in amorphous solids. Various properties, such as thermal
transport, force propagation, the vibrational spectrum and
resistance to flow present phenomena that are not satisfactorily
understood. An inherent difficulty in understanding these phenomena
is that they are often governed by processes that occur on small
length scales,  of the order of a few particle sizes.

The discovery that the jamming threshold of athermal idealized
spheres exhibits many properties of a critical point
\shortcite{epitome} changes this state of affairs, as it enables one
to separate the length scale on which disorder matters from the
molecular scale. This has led to the realization that the spectrum
of amorphous solids is characterized by a crossover, distinct from
Anderson localization, above which modes are extended but not
plane-wave-like, and that the characteristic length at which this
crossover occurs is decoupled from the molecular length. The
corresponding excitations, the anomalous modes, allow one to
understand several seemingly disparate anomalies in glasses and
granular materials in a coherent fashion.   The  picture advanced in
this review suggests that (1) the length scale $l^*$ corresponding
to the crossover characterizes force propagation in granular
packings~\shortcite{respprl,EllenbroekPRE09}, (2) the frequency
scale of the crossover corresponds to that of the boson peak in
glasses~\shortcite{silbertPRL05,wyartEPL,Xu07,wyartthesis}, (3) the
low-energy diffusivity of the modes above the crossover is
responsible for the mild linear increase of the thermal conductivity
with temperature above the plateau in
glasses~\shortcite{Xu09,Vitelli09,matthieucpap},
(4)  at least for colloidal glasses, the microscopic structure  is
marginally stable toward anomalous
modes~\shortcite{brito07a,brito3}, (5)
the modes appear to govern structural rearrangements in liquids at least in the viscosity range that can be probed numerically \shortcite{brito07b,brito3,harrowell}
 and
 (6) the structure of the modes corresponds to that of structural arrangements in amorphous solids 
 under mechanical load~\shortcite{xu09b,manning}.
At the linear level those excitations and their consequences are now
becoming rather well-understood theoretically for sphere packings,
with the notable exception of {\em (i) } the observed
quasi-localization of the lowest-frequency anomalous modes
\shortcite{xu09b,zeravcic09} and {\em (ii)} the apparently fractal
statistics of force propagation below the cut-off length $\ell^*$
above which continuum elasticity applies.

The jamming community is now beginning to study the non-linearities
coupling these excitations.  Much would be gained from such
knowledge. On the one hand,  it is possible that the
lowest-frequency anomalous modes are related to two-level systems,
which have been proposed on phenomenological grounds to explain
transport and specific heat anomalies at sub-Kelvin temperatures.
Despite decades of research, the spatial nature of two-level systems
has remained elusive.   If there is indeed a connection between the
highly anharmonic, low-frequency, quasilocalized modes and two-level
systems, then the energy barrier separating the two levels should
vanish as the isostatic jamming transition is approached.   On the
other hand, since the low-frequency modes are involved in
irreversible rearrangements, understanding their non-linearity is
presumably essential for a microscopic description of the flow of
supercooled liquids, granular matter and foams.

An exciting recent development is the advent of experiments capable
of studying the vibrational properties of systems near the jamming
transition~\shortcite{abatePRE06,bonn,Yodh,Gardel}.  In laboratory
systems, particle motion is typically damped and the inter-particle
potentials are not necessarily known; this has made it difficult to
make direct comparisons between experimental results and numerical
and theoretical predictions until now.  These complications have
recently been overcome~\shortcite{bonn,Yodh}, so experiments should
soon be able to test the extent to which predictions for ideal
spheres apply to real systems.

\section*{Acknowledgement}
We would like to thank Carolina Brito, Giulio Biroli, Jean-Philippe
Bouchaud, Ke Chen, Doug Durian, Wouter Ellenbroek, Jerry Gollub,
Peter Harrowell, Tom Haxton, Silke Henkes, Heinrich Jaeger,
Alexandre Kabla, Randy Kamien, Steve Langer, Haiyi Liang, Tom
Lubensky, L. Mahadevan, Xiaoming Mao, Kerstin Nordstrom, Corey
O'Hern, David Reichman, Leo Silbert, Anton Souslov, Brian Tighe,
Vincenco Vitelli, Martin van Hecke, Tom Witten, Erik Woldhuis, Ning
Xu, Arjun Yodh, Zorana Zeravcic, and Zexin Zhang for valuable
discussions, input, criticism and collaborations.

In addition, funding from DOE via DE-FG02-03ER46088 (SRN) and
DE-FG02-05ER46199 (AJL) as well as the NSF MRSEC program via
DMR-0820054 (SRN) and DMR05-20020 (AJL) is gratefully acknowledged.
AJL and SRN both thank the Kavli Institute for Theoretical Physics,
Santa Barbara, and AJL, SRN and MW thank the Aspen Center for
Physics for their hospitality. WvS would also like to thank in
particular Martin van Hecke for his help and support during the last
few years, and FOM for its generous support over the years.

\bibliographystyle{OUPnamed_notitle}
\bibliography{references}
%
\end{document}